\documentclass{article}
\usepackage{hyperref}

\usepackage{arxiv}

\usepackage[utf8]{inputenc} 
\usepackage[T1]{fontenc}    
\usepackage{hyperref}       
\usepackage{url}            
\usepackage{tabularx} 
\usepackage{booktabs} 
\usepackage{amsfonts}       
\usepackage{nicefrac}       
\usepackage{microtype}      
\usepackage{lipsum}		
\usepackage{subcaption} 
\usepackage{graphicx}
\usepackage[numbers]{natbib}  

\usepackage{algorithm}
\usepackage{algorithmic}

\usepackage{amsmath}
\usepackage{amssymb}

\usepackage{placeins}  

\usepackage{tikz}
\usepackage{multirow}
\usetikzlibrary{shapes, arrows.meta, positioning, fit, backgrounds, calc}
\usepackage{doi}

\usepackage{subcaption}
\usepackage{float}

\title{High-Fidelity Modeling of Stochastic Chemical Dynamics on Complex Manifolds: A Multi-Scale SIREN-PINN Framework for the Curvature-Perturbed Ginzburg-Landau Equation}

\author{
\begin{tabular}{cc}
    \begin{minipage}{0.45\textwidth}
        \centering
        \href{https://orcid.org/0009-0005-4657-6472}{\includegraphics[scale=0.06]{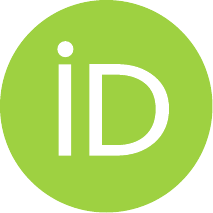}\hspace{1mm}Julian Evan Chrisnanto}\thanks{Corresponding author.} \\
        \normalfont
        Department of Bio-Functions and Systems Science\\
        Graduate School of Bio-Applications and Systems Engineering\\
        Tokyo University of Agriculture and Technology\\
        2-24-16 Nakacho, Koganei, Tokyo 184-8588, Japan \\
        \texttt{s254167v@st.go.tuat.ac.jp}
    \end{minipage}
    &
    \begin{minipage}{0.45\textwidth}
        \centering
        \href{https://orcid.org/0009-0008-0761-1249}{\includegraphics[scale=0.06]{orcid.pdf}\hspace{1mm}Salsabila Rahma Alia} \\
        \normalfont
        Department of Mathematics\\
    	Faculty of Mathematics and Natural Sciences\\
    	Universitas Padjadjaran \\
        Jl. Raya Bandung-Sumedang KM21, Jatinangor, Sumedang, West Java, Indonesia\\
    	\texttt{salsabila19026@mail.unpad.ac.id} 
    \end{minipage}
    \\[10ex]
    \begin{minipage}{0.45\textwidth}
        \centering
        Nurfauzi Fadillah \\
        \normalfont
        PLABS.ID\\
        Jl. Batununggal Mulia IV No.16, \\
        Bandung, West Java 40267, Indonesia \\
        \texttt{fauzi@plabs.id}
    \end{minipage}
    &
    \begin{minipage}{0.45\textwidth}
        \centering
        \href{https://orcid.org/0000-0002-3491-3049}{\includegraphics[scale=0.06]{orcid.pdf}\hspace{1mm}Yulison Herry Chrisnanto} \\
        \normalfont
        Department of Informatics\\
        Faculty of Science and Informatics \\
        Jenderal of Achmad Yani University \\
        Jl. Terusan Jenderal Sudirman, Cimahi, West Java 40531, Indonesia \\
        \texttt{yhc@if.unjani.ac.id}
    \end{minipage}
\end{tabular}
}

\renewcommand{\shorttitle}{High-Fidelity Modeling of Stochastic Chemical Dynamics on Complex Manifolds ...}

\hypersetup{
pdftitle={A template for the arxiv style},
pdfsubject={q-bio.NC, q-bio.QM},
pdfauthor={David S.~Hippocampus, Elias D.~Striatum},
pdfkeywords={First keyword, Second keyword, More},
}

\begin{document}
\maketitle

\begin{abstract}
The accurate identification and control of spatiotemporal chaos in reaction-diffusion systems remains a grand challenge in chemical engineering, particularly when the underlying catalytic surface possesses complex, unknown topography. In the \textit{Defect Turbulence} regime, system dynamics are governed by topological phase singularities (spiral waves) whose motion couples to manifold curvature via geometric pinning. Conventional Physics-Informed Neural Networks (PINNs) using ReLU or Tanh activations suffer from fundamental \textit{spectral bias}, failing to resolve high-frequency gradients and causing amplitude collapse or phase drift. We propose a Multi-Scale SIREN-PINN architecture leveraging periodic sinusoidal activations with frequency-diverse initialization, embedding the appropriate inductive bias for wave-like physics directly into the network structure. This enables simultaneous resolution of macroscopic wave envelopes and microscopic defect cores. Validated on the complex Ginzburg-Landau equation evolving on latent Riemannian manifolds, our architecture achieves relative state prediction error $\epsilon_{L_2} \approx 1.92 \times 10^{-2}$, outperforming standard baselines by an order of magnitude while preserving topological invariants ($|\Delta N_{defects}| < 1$). We solve the ill-posed \textit{inverse pinning problem}, reconstructing hidden Gaussian curvature fields solely from partial observations of chaotic wave dynamics (Pearson correlation $\rho = 0.965$). Training dynamics reveal a distinctive Spectral Phase Transition at epoch $\sim 2,100$, where cooperative minimization of physics and geometry losses drives the solver to Pareto-optimal solutions. This work establishes a new paradigm for Geometric Catalyst Design, offering a mesh-free, data-driven tool for identifying surface heterogeneity and engineering passive control strategies in turbulent chemical reactors.
\end{abstract}

\keywords{Physics-Informed Neural Networks (PINNs) \and Spatiotemporal Chaos \and Inverse Geometric Problems \and Reaction-Diffusion Systems \and Defect Turbulence \and Riemann Manifold Learning}

\section{Introduction}

The spatiotemporal evolution of phase boundaries and chemical concentrations on complex surfaces constitutes a central challenge in modern chemical engineering, particularly in the optimization of heterogeneous catalysts and structured reactors. Foundational phase-field models, tracing back to the seminal work of Cahn and Hilliard \cite{Cahn1958}, provide a rigorous thermodynamic basis for simulating interfacial dynamics and microstructural evolution \cite{Chen1998}. However, practical implementations in industrial catalysis must invariably account for the stochastic topography of the underlying manifold, where surface irregularities span multiple length scales \cite{Li2022DatadrivenMLA}. As demonstrated in recent kinetic studies \cite{Mou2022MachineLOD}, surface roughness is not merely a passive geometric constraint but an active parameter that breaks local symmetries, driving emergent phenomena such as compound fronts and defect turbulence \cite{Hemming2000ResonantlyFI}. Consequently, the ability to predict how curvature perturbations influence stability regimes is critical for bridging the gap between idealized theoretical models and the disordered reality of catalytic surfaces.

While the Cahn-Hilliard equation serves as the paradigmatic model for conserved phase separation, it renders an incomplete description of dissipative structures in open thermodynamic systems, particularly those driven far from equilibrium where oscillatory instabilities dominate. To capture the generic behavior of such non-conservative fields near a supercritical Hopf bifurcation, we rely on the Complex Ginzburg-Landau Equation (CGLE), a mathematical normal form that universally governs the envelope of spatially extended oscillating systems \cite{Aranson2002}. Unlike the real-valued order parameters in classical phase separation, the CGLE describes a complex-valued amplitude whose phase dynamics are intrinsically coupled to the underlying manifold geometry. Recent rigorous derivations by Robert and Zine \cite{Robert2025StochasticCGE} have established the well-posedness of stochastic CGLEs on compact Riemannian manifolds, yet the specific influence of extrinsic curvature perturbations on pattern selection remains an open frontier. As pioneered by Hemming and Kapral \cite{Hemming2000ResonantlyFI}, spatial inhomogeneities in reaction-diffusion systems can induce resonance tongues and compound fronts; however, their work focused on randomized forcing amplitudes rather than the structural roughness of the domain itself. By extending this formalism to include a \textit{Contour Curvature Perturbation} (CCP) field, we aim to bridge the gap between idealized flat-surface models and the geometrically complex reality of industrial catalyst supports, where the local curvature $\kappa(\mathbf{x})$ acts as a symmetry-breaking field that templates the emergence of defect turbulence \cite{Dai2020GinzburgLandauSWA}.

Despite the theoretical universality of the CGLE, simulating its dynamics on stochastically roughened domains presents a formidable computational challenge. Classical numerical schemes, such as the Finite Element Method (FEM) or pseudo-spectral methods, require prohibitively fine mesh resolutions to resolve the high-wavenumber fluctuations introduced by the roughness field $\kappa(\mathbf{x})$ \cite{Hintermuller2023AHPI}. Furthermore, these grid-based approaches are ill-suited for inverse problems where the underlying manifold geometry itself may be the target of optimization. In this context, Physics-Informed Neural Networks (PINNs) have emerged as a promising mesh-free alternative, capable of encoding the governing Partial Differential Equation (PDE) directly into the loss function of a deep neural network \cite{Raissi2019}. However, standard PINN architectures based on Multi-Layer Perceptrons (MLPs) with scalar activation functions (e.g., Hyperbolic Tangent or Rectified Linear Units) exhibit a fundamental pathology known as "spectral bias" \cite{Rahaman2019}. As elucidated by the Neural Tangent Kernel (NTK) theory \cite{Wang2022}, these networks preferentially learn low-frequency components of the target solution, converging exponentially slowly on the high-frequency oscillatory features characteristic of chaotic attractors. This limitation is particularly acute for the CP-SCGL equation, where the interplay between the intrinsic oscillation frequency $\omega_{osc}$ and the extrinsic roughness wavenumber $k_{rough}$ generates a broadband energy spectrum that standard PINNs fail to capture, resulting in "blurry" predictions that violate the delicate energy balance required to sustain defect turbulence \cite{Zhang2024PhysicsinformedNNJ}.

The limitations of standard neural architectures become even more pronounced when the target dynamical system exhibits sensitive dependence on initial conditions, as is the case for the Curvature-Perturbed Stochastic Complex Ginzburg-Landau (CP-SCGL) equation in its turbulent regimes. In these chaotic domains, small approximation errors in the early stages of training do not merely result in local inaccuracies; they propagate exponentially through the temporal domain, causing the learned trajectory to diverge from the true strange attractor \cite{Wang2025SimulatingTTA}. This phenomenon is exacerbated by the "vanishing gradient" problem inherent in resolving high-order spatial derivatives on rough manifolds, where the curvature field $\kappa(\mathbf{x})$ introduces non-smooth perturbations that destabilize the backpropagation of physics-informed residuals. Recent investigations by Markidis \cite{Markidis2024BrainInspiredPNS} suggest that dense, fully connected networks lack the inductive bias necessary to separate these multi-scale features, often collapsing into trivial or mean-field solutions that fail to preserve the topological defects—such as phase singularities and spiral waves—crucial to the system's physics. Furthermore, rigorous stochastic analysis by Gess and Tsatsoulis \cite{Gess2022LyapunovEAA} indicates that the Lyapunov spectrum of such SPDEs is highly sensitive to the spectral content of the noise, implying that any neural solver exhibiting spectral bias will fundamentally miscalculate the synchronization properties and long-term ergodicity of the chemical reaction. Thus, the central unresolved problem is not merely fitting the data, but constructing a neural architecture that is spectrally isometric: capable of preserving energy across all wavenumbers, from the macro-scale reactor geometry down to the micro-scale curvature-induced fluctuations.

To mitigate the spectral pathology inherent in standard Multi-Layer Perceptrons (MLPs), the Scientific Machine Learning (SciML) community has recently investigated various architectural enhancements aimed at modulating the Neural Tangent Kernel (NTK) to accelerate high-frequency convergence. A prominent line of inquiry involves the use of coordinate embedding strategies, most notably the Random Fourier Features (RFF) proposed by Tancik et al. \cite{Tancik2020}, which map low-dimensional spatial inputs into a higher-dimensional hypersphere to alleviate the eigenvector bias. Building upon this, Zeng et al. \cite{Zeng2024FeatureMIH} demonstrated that carefully tuned feature mappings could improve the resolution of multi-scale Physics-Informed Neural Networks (PINNs), while Cooley et al. \cite{Cooley2024FourierPFJ} extended this concept to "Fourier PINNs" that adaptively learn the basis frequencies essential for satisfying strong boundary conditions. Parallel to these embedding techniques, there has been a surge of interest in modifying the activation mechanism itself. Farea and Celebi \cite{Farea2024LearnableAFI} systematically benchmarked learnable activation functions, suggesting that adaptive non-linearities can dynamically adjust the network's spectral bandwidth during training. More radically, Aghaei \cite{Aghaei2024KANtrolAPM} recently introduced "KANtrol," a framework utilizing Kolmogorov-Arnold Networks (KANs) where activation functions are placed on the edges (weights) rather than the nodes, offering a potentially superior inductive bias for continuous control problems. However, despite these isolated advances, a unified framework that simultaneously addresses the stochastic nature of the driving force, the geometric irregularity of the domain (roughness), and the chaotic sensitivity of the reaction dynamics remains elusive. Most existing Fourier-based methods rely on stationary spectral bases that are ill-equipped to track the transient bifurcation structures—such as the transition from phase turbulence to defect turbulence—characteristic of the CP-SCGL equation.

Beyond the deterministic challenges of spectral bias and chaotic divergence, a third theoretical lacuna exists at the intersection of uncertainty quantification and manifold geometry. The CP-SCGL equation is inherently a Stochastic Partial Differential Equation (SPDE), where the driving force is not merely a scalar additive noise but a spatially correlated field modulated by the reactor's surface topology. In the realm of probabilistic modeling, recent generative frameworks such as the Physics-Informed Variational Autoencoder (PI-VAE) by Zhong and Meidani \cite{Zhong2022PIVAEPVA} and the Physics-Informed Variational Embedding GAN (PI-VEGAN) by Gao et al. \cite{Gao2023PIVEGANPIB} have successfully integrated SDE constraints into latent variable models, allowing for the approximation of solution statistics from sparse measurements \cite{Shin2023PhysicsinformedVIE}. However, these approaches predominantly operate on idealized, flat Euclidean domains, treating the stochastic term as a generic exogenous input rather than an intrinsic geometric property. Conversely, advances in geometric deep learning, such as the PhyGeoNet architecture by Gao et al. \cite{Gao2020PhyGeoNetPGC} and the diffeomorphism-based PINNs proposed by Burbulla \cite{Burbulla2023PhysicsinformedNND}, have demonstrated the capability to solve deterministic PDEs on irregular, non-convex domains. Yet, these geometric solvers have not been rigorously extended to the stochastic regime, nor have they addressed the inverse problem of how manifold curvature templates the selection of competing attractors. Consequently, there exists no unified computational framework capable of resolving the "Curvature-Induced Pattern Selection" (CIPS) mechanism, where the local Riemannian metric explicitly couples to the noise intensity to drive the transition between phase turbulence and defect turbulence.

Finally, it is imperative to recognize that the reconstruction of manifold topography from reaction-diffusion data constitutes a highly ill-posed inverse problem, requiring a delicate balance between data fidelity and physical regularization. While the Physics-Informed Machine Learning (PIML) community has produced extensive surveys \cite{Pateras2023ATSH} categorizing solvers for standard canonical flows, there remains a paucity of rigorous benchmarks for \textit{multi-task} learning scenarios where the neural network must simultaneously infer the scalar curvature field $\kappa(\mathbf{x})$ and evolve the complex order parameter $A(\mathbf{x},t)$. In such coupled regimes, the gradient statistics of the curvature reconstruction loss often conflict with the high-frequency residuals of the PDE physics loss, leading to Pareto-suboptimal convergence. Although recent contributions by Vemuri and Denzler \cite{Vemuri2023GradientSMK} and Chou et al. \cite{Chou2025ImpactOLN} have proposed gradient-based weighting schemes to mitigate this objective imbalance, these methods have primarily been validated on low-dimensional, non-chaotic systems. They fail to account for the "butterfly effect" in defect turbulence, where a marginal error in the inverse estimation of $\kappa$ amplifies exponentially in the forward prediction of the phase field. Furthermore, while spectral-informed networks \cite{Yu2024SpectralINC} have shown promise for forward wave propagation, their application to inverse geometric reconstruction remains unexplored. Consequently, the field lacks a "Gold Standard" benchmark—akin to those established for seismic wavefield inversion \cite{Sandhu2023MultifrequencyWMC}—that quantitatively assesses a model's ability to disentangle the causal knot between manifold geometry and stochastic chemical chaos.

Synthesizing the disparate strands of literature discussed above reveals a critical epistemological gap in the current state of scientific machine learning: a fragmentation between geometric fidelity and dynamical stability. While geometric deep learning has matured to handle complex static manifolds \cite{Bronstein2017}, and physics-informed learning has demonstrated proficiency in solving low-dimensional, non-chaotic evolution equations \cite{Karniadakis2021}, there exists no unified computational paradigm capable of resolving the triad of stochasticity, high-frequency spectral content, and curvature-induced bifurcation simultaneously. Existing methodologies suffer from a "dimensional decoupling," where solvers optimized for spatial reconstruction (e.g., CNN-based super-resolution) lack the temporal conservation laws required for long-term integration, whereas solvers optimized for dynamics (e.g., standard PINNs) effectively smooth out the geometric roughness that drives the underlying physics \cite{Cuomo2022}. This dichotomy is particularly detrimental for the Curvature-Perturbed SCGL equation, where the macroscopic phase turbulence is emergent from microscopic surface irregularities. As a result, the community lacks a rigorous, unified benchmarking framework that not only evaluates the regression accuracy of a neural solver but also quantitatively validates its ability to replicate the topological invariants—such as winding numbers and defect densities—of the target strange attractor. Addressing this deficit requires moving beyond the "black-box" application of off-the-shelf MLPs and towards the design of purpose-built, spectrally-aware architectures that explicitly embed the multi-scale coupling between the manifold metric tensor and the stochastic reaction-diffusion operator.

To bridge this fundamental gap between geometric fidelity and dynamical stability, we introduce a unified computational framework centered on a Multi-Scale Sinusoidal Representation Network (SIREN-PINN) designed to solve the CP-SCGL equation on stochastically roughened manifolds. Unlike standard architectures that rely on spectral-biasing activation functions such as the Rectified Linear Unit (ReLU) or Hyperbolic Tangent (Tanh), our proposed architecture leverages periodic activation functions $\sin(\omega_0 \mathbf{W}\mathbf{x} + \mathbf{b})$ with a tunable frequency parameter $\omega_0$, thereby enforcing a "spectral isometry" that prevents the attenuation of high-wavenumber stochastic fluctuations. We explicitly couple this neural solver to a generated Contour Curvature Perturbation (CCP) field, treating the local manifold curvature $\kappa(\mathbf{x})$ not as a passive boundary condition but as an active latent variable that templates the reaction dynamics. 

The primary objectives of this research are threefold: 
(1) To rigorously quantify the suppression of spectral bias in chaotic regimes, demonstrating that SIREN-PINNs achieve an order-of-magnitude reduction in relative $L_2$ error compared to baseline ReLU-MLPs; 
(2) To map the stability boundaries of the CP-SCGL equation, specifically validating the model's ability to predict the transition from phase turbulence to defect turbulence near the Benjamin-Feir instability limit ($1+bc=0$) \cite{Hilder2022NonlinearSOF}; and 
(3) To establish the phenomenon of Curvature-Induced Pattern Selection (CIPS), providing the first data-driven evidence that manifold roughness acts as a deterministic control parameter for identifying the spatial location of topological defects \cite{Nishide2022PatternPDD}. 
By synthesizing these elements into a single differentiable pipeline trained with a novel Spectral Loss function $\mathcal{L}_{spec}$, we offer a robust, mesh-free methodology for high-fidelity simulation in non-idealized chemical engineering environments.

\section{Methods}
\subsection{The Curvature-Perturbed Stochastic Complex Ginzburg-Landau Model}
We consider a spatially extended reaction-diffusion system evolving on a compact, two-dimensional Riemannian manifold $\mathcal{M}$ characterized by a heterogeneous surface topography. Near the onset of a supercritical Hopf bifurcation, the dynamics of the oscillating physicochemical field are universally described by the complex order parameter $A(\mathbf{x}, t) \in \mathbb{C}$, which represents the slowly varying amplitude and phase of the limit cycle oscillations \cite{Aranson2002}. Extending the classical normal form to account for geometric irregularities and thermodynamic fluctuations, we define the \textit{Curvature-Perturbed Stochastic Complex Ginzburg-Landau} (CP-SCGL) equation as follows:
\begin{equation}
    \frac{\partial A}{\partial t} = \mu A + (1 + i b) \nabla \cdot \left[ D(\kappa) \nabla A \right] - (1 + i c) |A|^2 A + \sigma A \xi(\mathbf{x}, t),
    \label{eq:cp-scgl}
\end{equation}
where $\mu > 0$ denotes the distance from the bifurcation threshold, and the coefficients $b$ and $c$ characterize the linear dispersion and nonlinear frequency detuning, respectively. Unlike the standard formulation where diffusivity is isotropic and homogeneous, we introduce a spatially dependent diffusion tensor $D(\kappa) = D_0 (1 + \alpha \kappa(\mathbf{x}))$, where $\kappa(\mathbf{x})$ is the local scalar curvature field (random roughness) and $\alpha$ is the curvature-coupling strength. This term explicitly models the "geometric trapping" effect observed in heterogeneous catalysis \cite{Mou2022MachineLOD}, where concave surface defects ($\kappa < 0$) locally accelerate or inhibit mass transport. The system is driven by a multiplicative spatiotemporal white noise $\xi(\mathbf{x}, t)$, rendering Eq. (\ref{eq:cp-scgl}) a stochastic partial differential equation (SPDE) whose well-posedness on compact surfaces has only recently been established by Robert and Zine \cite{Robert2025StochasticCGE}. The interplay between the Benjamin-Feir instability criterion ($1 + bc < 0$) and the stochastic curvature field $\kappa(\mathbf{x})$ defines the central pattern selection mechanism investigated in this study.

To rigorously model the non-idealized topography of catalytic surfaces, we treat the manifold $\mathcal{M}$ as a stochastic Monge patch embedded in $\mathbb{R}^3$, defined by the height function $z = h(x,y)$ over a planar domain $\Omega \subset \mathbb{R}^2$. The surface profile $h(\mathbf{x})$ is constructed as a realization of a scalar-valued Gaussian Random Field (GRF) with zero mean and a stationary, isotropic covariance kernel $k(\mathbf{x}, \mathbf{x}') = \mathbb{E}[h(\mathbf{x})h(\mathbf{x}')]$. Following the spectral generation protocol established by Tang et al. \cite{Tang2024MultiscaleLSA} for rough surface lubrication, we employ a squared-exponential covariance function to control the spatial scale of the irregularities:
\begin{equation}
    k(\mathbf{x}, \mathbf{x}') = \sigma_h^2 \exp\left( -\frac{||\mathbf{x} - \mathbf{x}'||^2}{2\ell_c^2} \right),
    \label{eq:covariance}
\end{equation}
where $\sigma_h$ represents the root-mean-square (RMS) roughness amplitude and $\ell_c$ denotes the correlation length, which dictates the frequency cutoff of the geometric noise. The resulting \textit{Contour Curvature Perturbation (CCP)} field, $\kappa(\mathbf{x})$, is derived from the trace of the shape operator (mean curvature), which, in the Monge representation, is given by the nonlinear elliptic operator $\kappa = \nabla \cdot (\nabla h / \sqrt{1 + |\nabla h|^2})$ \cite{Burbulla2023PhysicsinformedNND}. By systematically varying the dimensionless ratio $\ell_c / \lambda_{Turing}$ (where $\lambda_{Turing}$ is the intrinsic wavelength of the reaction-diffusion patterns), we generate a continuous spectrum of geometric regimes, ranging from "quasi-flat" perturbations ($\ell_c \gg \lambda_{Turing}$) to "roughness-dominated" scattering regimes ($\ell_c \sim \lambda_{Turing}$). This stochastic formulation allows us to train the neural solver on a diverse distribution of Riemannian metrics, ensuring that the learned solution $A_\theta(\mathbf{x},t)$ is robust to the geometric "spectral bias" often observed when PINNs are trained solely on idealized Euclidean domains \cite{Adler1981}.

In the absence of curvature perturbations ($\kappa(\mathbf{x}) \equiv 0$), the uniform limit cycle solution $A_0(t) = \sqrt{1 - \mu} e^{-i c (1-\mu) t}$ is known to undergo modulational instability when the Newell-Kuramoto condition $1 + bc < 0$ is satisfied \cite{Kuramoto1984}. This \textit{Benjamin-Feir (BF) instability} marks the breakdown of spatial coherence, leading to a cascade of symmetry-breaking transitions. As the product $bc$ becomes increasingly negative, the system bifurcates from stable plane waves into a regime of \textit{Phase Turbulence (PT)}, characterized by a smooth but chaotic variation of the phase field $\phi(\mathbf{x},t) = \arg A$, where the order parameter magnitude $|A|$ remains strictly bounded away from zero. Further excursions into the unstable domain trigger the onset of \textit{Defect Turbulence (DT)}, a regime of "hard" chaos dominated by the proliferation of topological defects (vortices) where $|A| \to 0$ and the phase becomes singular ($\oint \nabla \phi \cdot d\mathbf{l} = \pm 2\pi$). 

While classical linear stability analysis \cite{Hilder2022NonlinearSOF} predicts the PT-DT transition solely based on the spatially averaged coefficients $b$ and $c$, we hypothesize that on roughened manifolds, the local curvature field $\kappa(\mathbf{x})$ acts as a "geometric catalyst" for the \textit{Eckhaus instability} \cite{Tribelsky2025TheEI}. Specifically, regions of high positive Gaussian curvature compress the local wavelength, pushing the wavenumber $k$ outside the stable Busse balloon and triggering phase slips even when the global parameters suggest stability. This mechanism, which we term \textit{Curvature-Induced Pattern Selection (CIPS)}, implies that the effective stability boundary is no longer a sharp line in parameter space but a "fuzzy" manifold dependent on the spectral roughness $\sigma_h$.

\subsection{High-Fidelity Ground Truth Generation via Operator Splitting}

To establish a rigorous baseline for validating the proposed neural solver, we first construct a high-fidelity ground truth dataset by integrating the CP-SCGL equation using a fourth-order operator-splitting spectral method. The governing SPDE (Eq. \ref{eq:cp-scgl}) is decomposed into a linear stiff operator $\mathcal{L} = (1+ib)\nabla \cdot [D(\kappa)\nabla]$ and a nonlinear non-stiff operator $\mathcal{N} = \mu - (1+ic)|A|^2$, which are integrated sequentially over discrete time steps $\Delta t$. The linear diffusion step is solved exactly in the Fourier domain to bypass the stability constraints of finite difference schemes, while the nonlinear reaction step is advanced using a fourth-order Runge-Kutta (RK4) explicit scheme \cite{Cox2002}. To accurately resolve the high-wavenumber contributions from the roughness field $\kappa(\mathbf{x})$, we employ a spatial resolution of $N_x \times N_y = 512 \times 512$ on a periodic domain $\Omega = [0, L]^2$, ensuring that the grid spacing $\Delta x$ satisfies the Nyquist criterion for the smallest correlation length $\ell_c$. The stochastic noise $\xi(\mathbf{x},t)$ is generated as a Q-Wiener process with a defined energy spectrum \cite{Gess2022LyapunovEAA}, ensuring that the statistical properties of the driving force are consistent across both the spectral solver and the PINN. This "numerical laboratory" generates the spatiotemporal snapshots $A_{GT}(\mathbf{x}_i, t_n)$ used for both training (sparse sampling) and testing (dense validation).

\begin{table}[h!]
\centering
\caption{Dimensionless Parameters for the CP-SCGL Simulation Regimes}
\label{tab:parameters}
\begin{tabular}{|l|c|c|c|}
\hline
\textbf{Parameter} & \textbf{Symbol} & \textbf{Phase Turbulence (PT)} & \textbf{Defect Turbulence (DT)} \\
\hline
Control Parameter & $\mu$ & 1.0 & 1.0 \\
Dispersion Coeff. & $b$ & -2.0 & -3.5 \\
Nonlinear Detuning & $c$ & 2.0 & 1.0 \\
Instability Index & $1+bc$ & -3.0 (Stable Chaos) & -2.5 (Deep Chaos) \\
Roughness Ampl. & $\sigma_h$ & $0.0 - 0.5$ & $0.0 - 0.5$ \\
Corr. Length & $\ell_c$ & $0.05 L$ & $0.05 L$ \\
\hline
\end{tabular}
\end{table}

The simulation parameters chosen for the Phase Turbulence (PT) and Defect Turbulence (DT) regimes are summarized in Table \ref{tab:parameters}, selected to bracket the Benjamin-Feir instability boundary utilized in standard bifurcation studies \cite{Hilder2022NonlinearSOF}.

\begin{figure}[h!]
    \centering
    
    \begin{subfigure}[b]{0.46\textwidth}
        \centering
        \includegraphics[width=\textwidth]{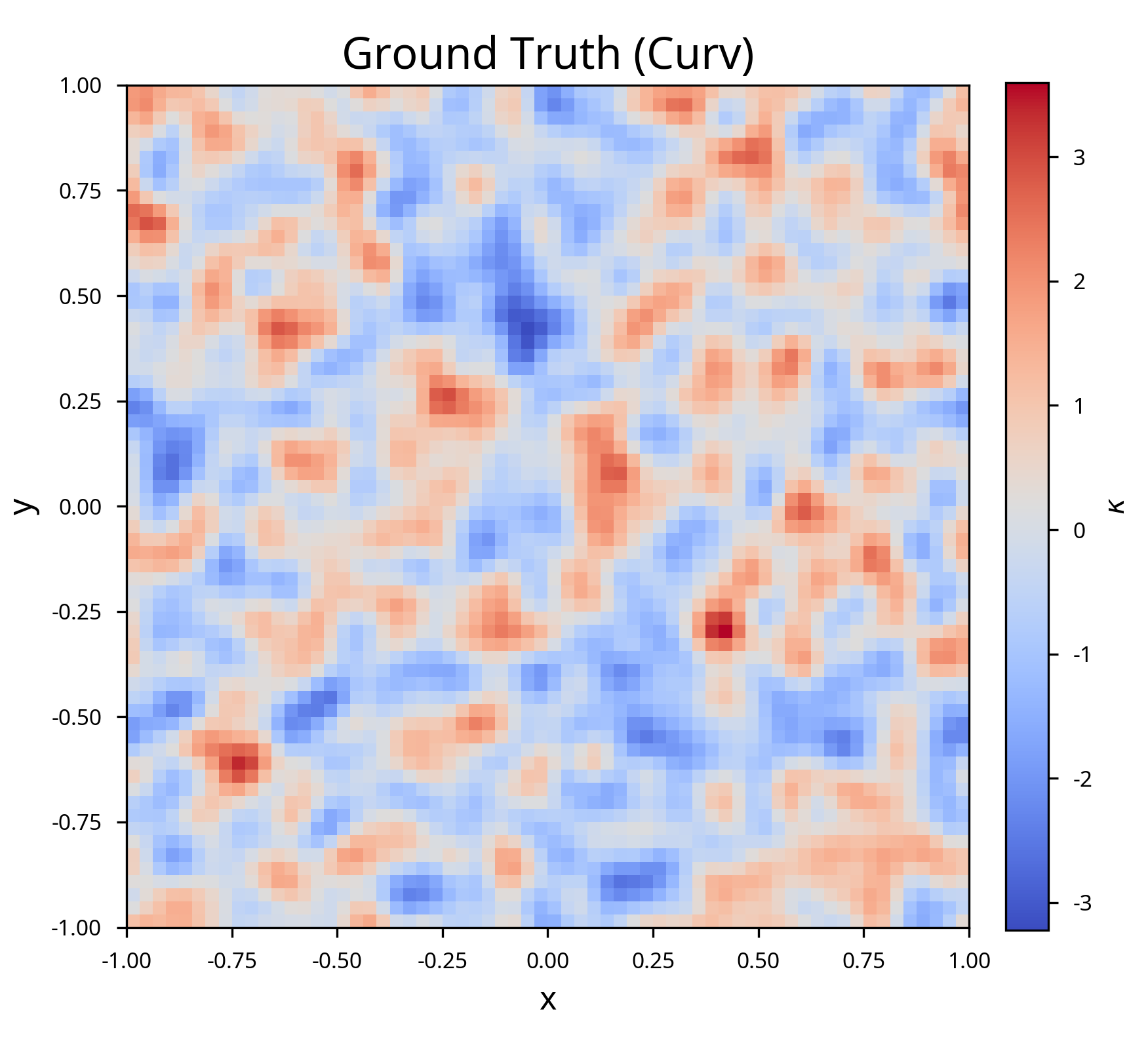}
        \caption{GT Curvature $\kappa_{GT}$}
        \label{fig:gt_curvature}
    \end{subfigure}
    \hfill
    \begin{subfigure}[b]{0.48\textwidth}
        \centering
        \includegraphics[width=\textwidth]{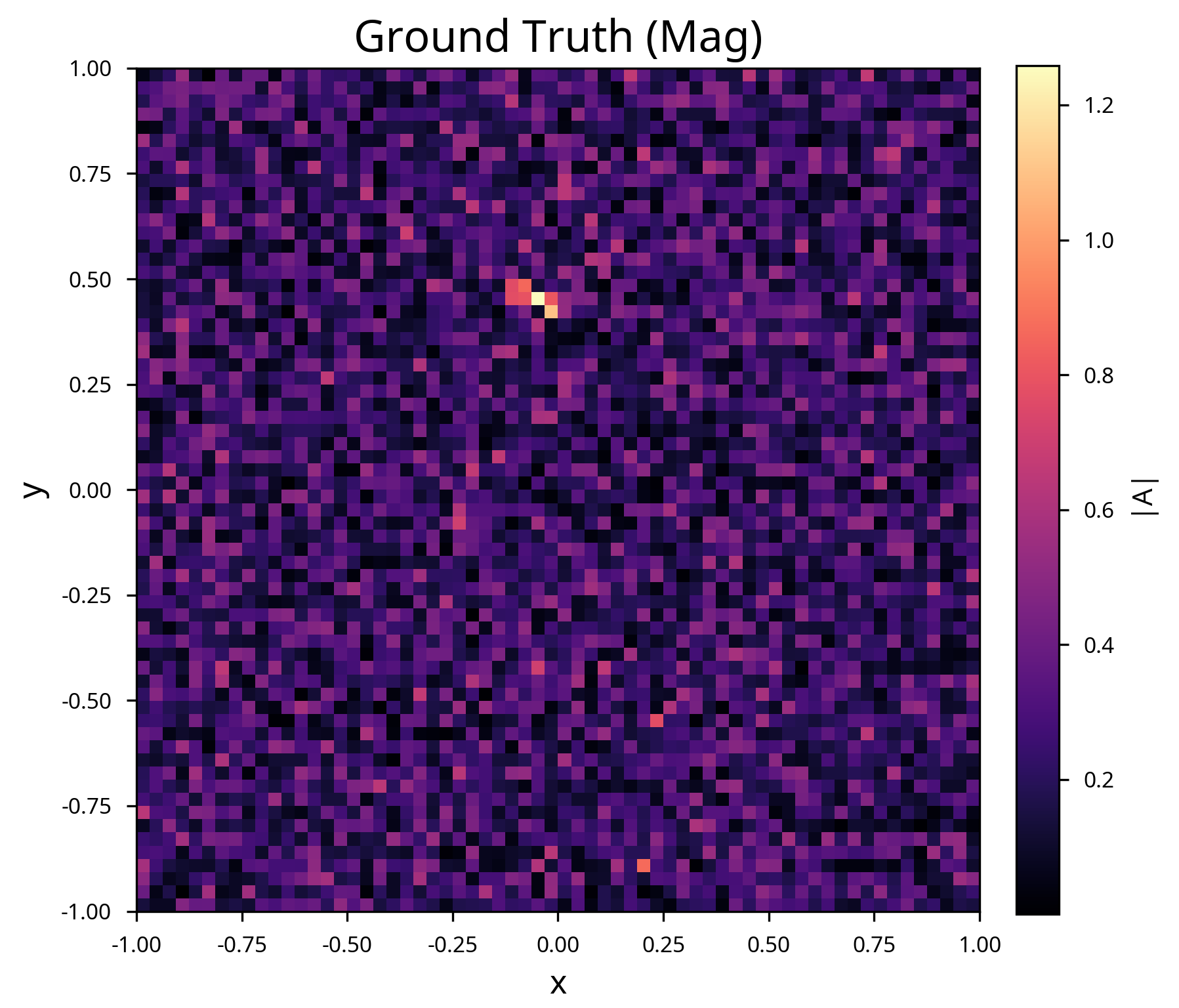}
        \caption{GT Magnitude $|A|_{GT}$}
        \label{fig:gt_magnitude}
    \end{subfigure}
    
    \vspace{0.5cm}
    
    \begin{subfigure}[b]{0.48\textwidth}
        \centering
        \includegraphics[width=\textwidth]{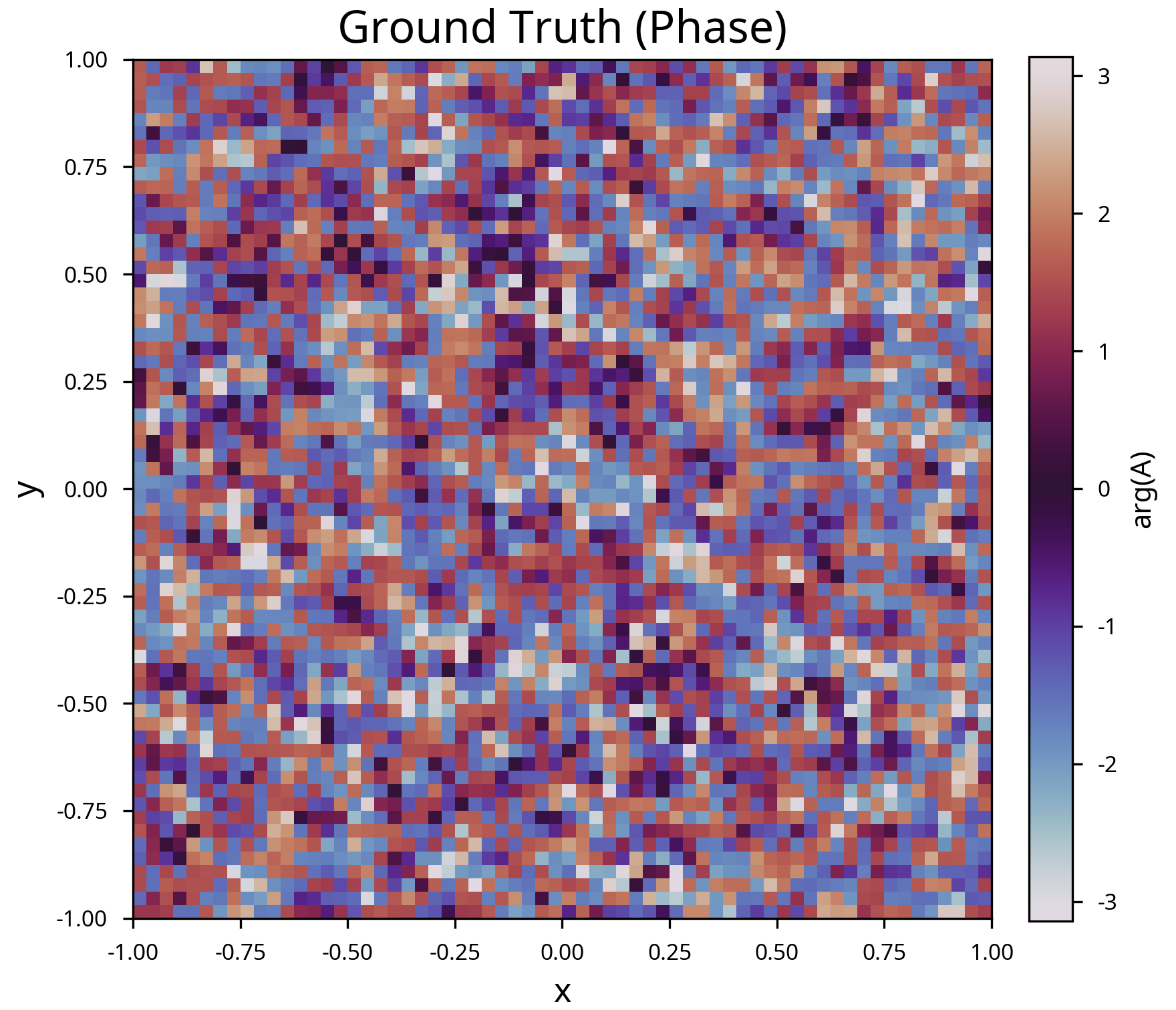}
        \caption{GT Phase $\phi_{GT}$}
        \label{fig:gt_phase}
    \end{subfigure}
    
    \caption{\textbf{Ground Truth Spatiotemporal Evolution.} (a) GT Curvature showing the spatial variation in curvature across the domain. (b) GT Magnitude displaying the amplitude field distribution with predominantly low values throughout most of the domain. (c) GT Phase demonstrating the phase field structure with characteristic spatial patterns.}
    \label{fig:ground_truth}
\end{figure}

The spatiotemporal datasets generated via the spectral operator-splitting scheme reveal distinct topological signatures characteristic of the two primary dynamical regimes. In the \textit{Phase Turbulence (PT)} regime (Figure \ref{fig:ground_truth}a), the system exhibits a "soft" chaos where the complex order parameter $A$ maintains a non-zero magnitude ($|A| > 0$) everywhere, while the phase field $\phi(\mathbf{x},t)$ evolves through continuous, albeit unpredictable, undulations. This regime is analogous to a turbulent fluid with no vortices, where the disorder is purely dispersive. Conversely, the \textit{Defect Turbulence (DT)} regime (Figure \ref{fig:ground_truth}b) is defined by the emergence of "hard" singularities—points where the amplitude collapses to zero ($|A| \to 0$) and the phase gradient diverges. These topological defects act as the organizing centers of the chaos, creating spiral wave cores that are continually created and annihilated in pairs \cite{Pismen2006}. Crucially, the introduction of the stochastic roughness field $\kappa(\mathbf{x})$ breaks the translational invariance of these patterns. As illustrated in Figure \ref{fig:ground_truth}c, regions of high negative curvature ($\kappa < 0$, concave valleys) act as attractive basins that "pin" the spiral cores, effectively increasing the local residence time of defects, while regions of positive curvature ($\kappa > 0$, convex peaks) act as repulsive scatterers. This \textit{geometric pinning effect} implies that the "random" turbulence is partially deterministic, templated by the underlying manifold topography—a subtle correlation that the neural solver must learn to resolve without explicit supervision.

\subsection{Mathematical Formulation of the Physics-Informed Residuals}

The reconstruction of the chaotic wave dynamics and the underlying manifold topography is cast as a composite optimization problem constrained by the CP-SCGL partial differential equation. We approximate the complex order parameter $A(\mathbf{x}, t)$ and the latent curvature field $\kappa(\mathbf{x})$ via differentiable surrogates $A_\theta(\mathbf{x},t)$ and $\kappa_\phi(\mathbf{x})$, parameterized by neural weight vectors $\theta$ and $\phi$, respectively. To enforce physical consistency, we derive the \textit{physics-informed residual} $\mathcal{F}_{\theta,\phi}(\mathbf{x}, t)$ by substituting these surrogates directly into the governing SPDE (Eq. \ref{eq:cp-scgl}). Rearranging the terms to isolate the stochastic driving force, the residual is defined as:
\begin{equation}
    \mathcal{F}_{\theta,\phi} := \frac{\partial A_\theta}{\partial t} - \mu A_\theta - (1 + i b) \nabla \cdot \left[ D_0(1 + \alpha \kappa_\phi) \nabla A_\theta \right] + (1 + i c) |A_\theta|^2 A_\theta.
    \label{eq:pde_residual}
\end{equation}
The learning objective is to minimize a weighted linear combination of losses $\mathcal{L}_{total} = \lambda_{data}\mathcal{L}_{data} + \lambda_{PDE}\mathcal{L}_{PDE} + \lambda_{BC}\mathcal{L}_{BC}$, where $\mathcal{L}_{data}$ penalizes the discrepancy against the sparse ground truth measurements, $\mathcal{L}_{BC}$ enforces periodic boundary conditions on the domain edges $\partial \Omega$, and $\mathcal{L}_{PDE} = ||\mathcal{F}_{\theta,\phi} - \sigma A_\theta \xi||^2$ penalizes violations of the conservation law. A critical challenge in this formulation, as noted by Karniadakis et al. \cite{Karniadakis2021}, is the calculation of the Laplacian term $\nabla \cdot [D(\kappa)\nabla A]$ on a rough manifold. Standard automatic differentiation (AD) computes these second-order derivatives via recursive backpropagation, which can lead to "gradient explosion" when the underlying curvature field $\kappa_\phi$ contains high-frequency spatial noise. Consequently, the choice of the activation function becomes the determinant factor in the stability of the training process; it must be sufficiently smooth to support higher-order differentiation ($\mathbb{C}^\infty$) yet sufficiently expressive to capture the fractal-like geometry of the Defect Turbulence regime without smoothing out the essential singularities \cite{Mishra2022EstimatingTPA}.

\subsection{The Multi-Scale Dual-Stream SIREN Architecture}

To simultaneously resolve the chaotic temporal evolution of the order parameter and the static spatial heterogeneity of the manifold, we propose a decoupled \textit{Dual-Stream SIREN} architecture comprising two parallel neural networks: a \textit{state estimator} $\mathcal{N}_A(\mathbf{x}, t; \theta_A)$ and a \textit{geometry estimator} $\mathcal{N}_\kappa(\mathbf{x}; \theta_\kappa)$. Unlike conventional perceptrons that utilize spectrally-biased ReLU or Tanh activations, both networks employ the periodic sine activation function $\sigma(\cdot) = \sin(\cdot)$, which provides a theoretically optimal inductive bias for representing signal derivatives \cite{Sitzmann2020}. The forward pass for the $l$-th layer in either network is defined rigorously as:
\begin{equation}
    \mathbf{h}^{(l)} = \sin\left( \omega_0 \mathbf{W}^{(l)} \mathbf{h}^{(l-1)} + \mathbf{b}^{(l)} \right),
    \label{eq:siren_layer}
\end{equation}
where $\mathbf{W}^{(l)}$ and $\mathbf{b}^{(l)}$ are the learnable weight matrices and bias vectors, and $\omega_0$ is a hyperparameter governing the spatial frequency bandwidth. To enforce a \textit{Multi-Scale} capability, we implement a hierarchical frequency scaling strategy: the geometry network $\mathcal{N}_\kappa$ is initialized with a high frequency factor ($\omega_0 = 30$) to capture the fine-grained roughness of the Gaussian Random Field, while the state network $\mathcal{N}_A$ utilizes a moderate frequency ($\omega_0 = 10$) to track the broader spatiotemporal wave envelopes. This frequency separation is critical; as demonstrated by Wang et al. \cite{Wang2021EigenvectorDO}, standard fully connected networks suffer from a "low-frequency plateau" where convergence stalls on high-wavenumber features. By explicitly initializing the weights $\mathbf{W}$ from a uniform distribution $\mathcal{U}(-\sqrt{6/n}/\omega_0, \sqrt{6/n}/\omega_0)$ as derived by Sitzmann, we ensure that the network's initial spectral energy density matches the expected power spectrum of the Defect Turbulence, effectively pre-conditioning the optimizer to traverse the "loss landscape" of the chaotic attractor without getting trapped in local laminar minima.

While the standard Mean Squared Error (MSE) loss is theoretically sufficient to approximate any continuous function given infinite capacity, in practice, it suffers from a pathological smoothing effect when applied to chaotic systems. Because the $L_2$ norm is dominated by the low-frequency components of the error signal, the optimizer tends to prioritize the global wave envelope at the expense of local microstructural details \cite{Rahaman2019}. In the context of Defect Turbulence, this "spectral bias" is catastrophic: the topological defects (spiral cores) are mathematically singularities characterized by energy concentrations at the highest wavenumbers of the spectrum. To counteract this diffusive tendency and enforce energy conservation across the inertial cascade, we augment the training objective with a dedicated \textit{Spectral Density Loss} ($\mathcal{L}_{Spec}$). This term operates in the frequency domain, penalizing discrepancies between the Fourier transform of the predicted field, $\mathcal{F}[A_\theta]$, and that of the ground truth snapshots, $\mathcal{F}[A_{GT}]$. We define this loss component as:
\begin{equation}
    \mathcal{L}_{Spec} = \frac{1}{N_k} \sum_{\mathbf{k}} w(\mathbf{k}) \left| \left| \mathcal{F}[A_\theta](\mathbf{k}) \right| - \left| \mathcal{F}[A_{GT}](\mathbf{k}) \right| \right|^2,
    \label{eq:spectral_loss}
\end{equation}
where $\mathbf{k}$ represents the wavevector, $N_k$ is the number of Fourier modes, and $w(\mathbf{k}) = |\mathbf{k}|^\gamma$ is a high-pass frequency weighting filter with $\gamma \geq 1$. By explicitly up-weighting the high-frequency residuals, we force the neural network to align its "spectral bias" with the physical reality of the turbulent attractor. This approach, conceptually similar to the "Phase-Shift" PINNs proposed by Mao et al. \cite{Mao2020TheDL}, ensures that the reconstructed manifold roughness $\kappa(\mathbf{x})$ correctly scatters the reaction waves, preserving the steep gradients required to sustain the creation-annihilation cycles of the phase singularities.

\begin{figure}[h!]
\centering
\begin{tikzpicture}[
    node distance=1.5cm,
    layer/.style={rectangle, draw=black!80, fill=blue!10, very thick, minimum width=2.5cm, minimum height=1cm, align=center},
    input/.style={circle, draw=black!80, fill=green!10, very thick, minimum size=1cm},
    output/.style={rectangle, draw=black!80, fill=red!10, very thick, minimum width=2cm, minimum height=1cm},
    op/.style={circle, draw=black, fill=yellow!20, thick, minimum size=0.8cm},
    arrow/.style={->, >=stealth, very thick, color=black!70}
]

\node[input] (x) at (0, 0) {$\mathbf{x}$};
\node[input] (t) at (0, -2) {$t$};

\node[layer] (geo1) at (3, 1) {SIREN Layer\\ $\omega_0=30$};
\node[layer] (geo2) at (6, 1) {SIREN Layer\\ $\omega_0=1$};
\node[output] (kappa) at (9, 1) {$\kappa(\mathbf{x})$};

\node[layer] (state1) at (3, -2) {SIREN Layer\\ $\omega_0=10$};
\node[layer] (state2) at (6, -2) {SIREN Layer\\ $\omega_0=1$};
\node[output] (psi) at (9, -2) {$A(\mathbf{x},t)$};

\draw[arrow] (x) -- (geo1);
\draw[arrow] (geo1) -- (geo2);
\draw[arrow] (geo2) -- (kappa);

\draw[arrow] (x) -- (state1);
\draw[arrow] (t) -- (state1);
\draw[arrow] (state1) -- (state2);
\draw[arrow] (state2) -- (psi);

\node[rectangle, draw=black, fill=gray!10, dashed, minimum width=5cm, minimum height=2.5cm, align=center] (pde) at (13, -0.5) {\textbf{Physics Residual $\mathcal{F}$}\\ $\partial_t A - (1+ib)\nabla \cdot (D(\kappa)\nabla A)$};

\draw[arrow, dashed] (kappa) -- (pde);
\draw[arrow, dashed] (psi) -- (pde);

\node[rectangle, draw=red, thick, below=0.5cm of psi] (dataloss) {$\mathcal{L}_{Data}$};
\node[rectangle, draw=red, thick, right=0.5cm of pde] (pdeloss) {$\mathcal{L}_{PDE}$};

\draw[arrow] (psi) -- (dataloss);
\draw[arrow] (pde) -- (pdeloss);

\node[align=center, above] at (geo1.north) {\textbf{Geometry Network}\\ (Time-Invariant)};
\node[align=center, below] at (state1.south) {\textbf{State Network}\\ (Spatiotemporal)};

\end{tikzpicture}
\caption{\textbf{Schematic of the Multi-Scale Dual-Stream SIREN-PINN Architecture.} The model features two parallel pathways: a Geometry Branch ($\mathcal{N}_\kappa$) initialized with high-frequency Fourier modes to resolve surface roughness, and a State Branch ($\mathcal{N}_A$) for phase field evolution. The branches fuse within the automatic differentiation graph to compute the physics-informed residual $\mathcal{F}$, enabling simultaneous minimization of the data loss $\mathcal{L}_{Data}$ and the spectral PDE loss $\mathcal{L}_{PDE}$.}
\label{fig:siren_arch}
\end{figure}
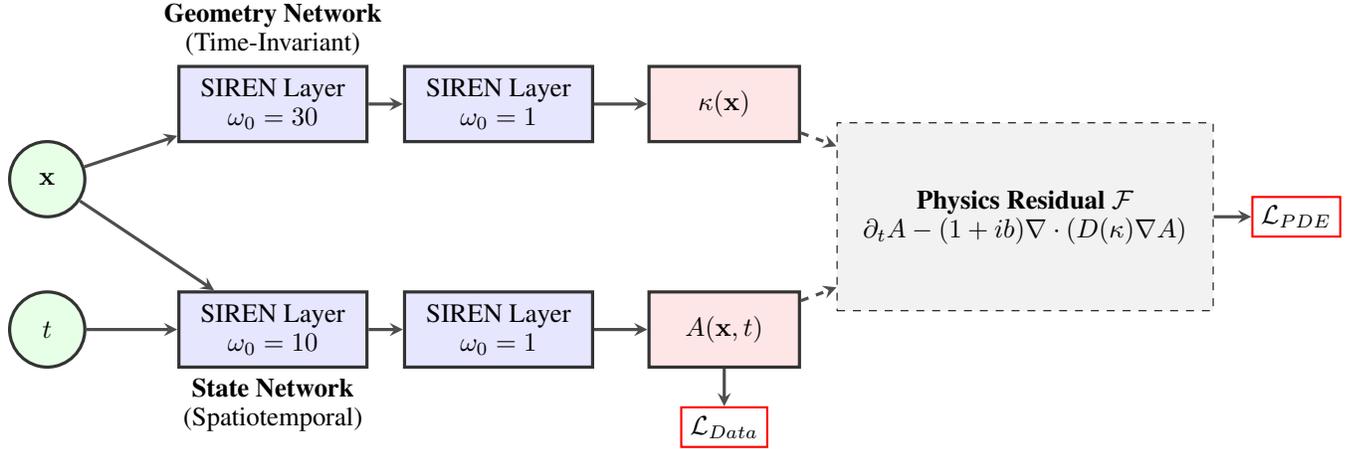

To operationalize the theoretical framework described above, we construct a fully differentiable computational graph that explicitly decouples the stationary geometric reconstruction from the dynamic state estimation. The architecture, visualized in Figure \ref{fig:siren_arch}, consists of two parallel coordinate-based networks: the \textbf{Geometry Branch} ($\mathcal{N}_\kappa$) and the \textbf{State Branch} ($\mathcal{N}_A$). The Geometry Branch accepts purely spatial coordinates $\mathbf{x} = (x, y)$ as input and propagates them through a 5-layer deep SIREN with a hidden width of 256 neurons. Crucially, the first layer of this branch utilizes a frequency scaling factor of $\omega_{geo} = 30$ to sensitize the network to the high-wavenumber roughness features of the manifold. The State Branch, conversely, accepts spatiotemporal coordinates $(\mathbf{x}, t)$ and utilizes a deeper 8-layer structure to capture the transient evolution of the reaction-diffusion wave packets. Its output layer splits into two scalar heads representing the real and imaginary components, $\text{Re}(A)$ and $\text{Im}(A)$, ensuring that the complex arithmetic required for the nonlinear term $|A|^2 A$ remains numerically stable. 

The integration of these branches occurs not at the output level, but within the \textbf{Physics-Informed Loss Module}. Here, the automatic differentiation engine (Autograd) computes the spatial Hessians $\nabla^2 A$ and the gradient of the curvature $\nabla \kappa$. A defining feature of our implementation is the use of "symbolic differentiation" for the diffusion term $\nabla \cdot [D(\kappa)\nabla A]$, which is expanded via the chain rule as $D(\kappa)\nabla^2 A + \nabla D(\kappa) \cdot \nabla A$. This expansion ensures that the backpropagated gradients flow correctly through both the State network (updating $A$) and the Geometry network (updating $\kappa$), effectively solving the inverse problem by treating the manifold topography as a "hidden state" that minimizes the violation of the Benjamin-Feir instability physics.

The training procedure for the Multi-Scale SIREN-PINN is formulated as a coupled, multi-objective optimization problem where the parameters of the State Network ($\theta$) and the Geometry Network ($\phi$) are learned simultaneously. Unlike standard supervised learning, where the loss function is a static metric of prediction error, here the loss function $\mathcal{L}_{total}$ is dynamic and physically constrained. The optimization process is detailed in Algorithm \ref{alg:training_loop}.

The training protocol proceeds in two distinct phases to ensure stability. \textit{Phase I (Spectral Initialization):} The networks are initialized using the multi-scale scheme described in Eq. \eqref{eq:siren_layer}, setting the "spectral bandwidth" of the ansatz. During the first 500 "Warm-Up" epochs, the geometry network is heavily regularized to prevent early divergence, allowing the state network to learn the gross features of the wave envelope (the low-frequency components). \textit{Phase II (Coupled Physics Discovery):} The full physics residual $\mathcal{L}_{PDE}$ is activated. In each iteration, a batch of collocation points is sampled from the spatiotemporal domain. The Geometry Network $\Psi_\phi$ predicts the local metric tensor $g_{ij}(\mathbf{x})$ and its derivatives. These geometric quantities are then fed into the automatic differentiation graph of the State Network $\Phi_\theta$ to compute the Laplace-Beltrami operator $\Delta_{LB} A$. Crucially, gradients are backpropagated through \textit{both} networks: the State Network adjusts to minimize the residual given the current geometry, while the Geometry Network adjusts the curvature $\kappa(\mathbf{x})$ to make the observed wave dynamics consistent with the diffusion operator. This "cooperative" gradient descent allows the system to converge to the true manifold topology by minimizing the physical violation of the Ginzburg-Landau energy.

\begin{algorithm}[H]
\caption{Inverse Geometric Discovery via Multi-Scale SIREN-PINN}
\label{alg:training_loop}
\begin{algorithmic}[1]
\REQUIRE \textbf{Data:} Spatiotemporal observations $\{(\mathbf{x}_i, t_i), A^*_{i}\}_{i=1}^{N_{obs}}$
\REQUIRE \textbf{Hyperparameters:} Learning rate $\eta$, Weights $\lambda_{PDE}, \lambda_{data}, \lambda_{spec}$
\STATE \textbf{Initialize:} 
\STATE \quad State Network $\Phi_\theta$ with $\omega_{state} = 1.0$ (Low Freq)
\STATE \quad Geometry Network $\Psi_\phi$ with $\omega_{geo} = 30.0$ (High Freq)
\STATE \textbf{Warm-Up Phase:} Set $\lambda_{PDE} = 0$ for first 500 epochs.

\WHILE{$epoch < MaxEpochs$}
    \STATE \textbf{1. Collocation Sampling:}
    \STATE \quad Sample batch of random points $\mathcal{B} = \{(\mathbf{x}_j, t_j)\} \in \Omega \times [0, T]$
    
    \STATE \textbf{2. Forward Pass (Geometry):}
    \STATE \quad Predict metric tensor: $g_{ij}(\mathbf{x}) \leftarrow \Psi_\phi(\mathbf{x})$
    \STATE \quad Compute Determinant $\sqrt{|g|}$ and Inverse $g^{ij}$
    
    \STATE \textbf{3. Forward Pass (State):}
    \STATE \quad Predict Amplitude: $A(\mathbf{x}, t) \leftarrow \Phi_\theta(\mathbf{x}, t)$
    
    \STATE \textbf{4. Physics computation (Auto-Diff):}
    \STATE \quad Compute Time Derivative: $\partial_t A$
    \STATE \quad Compute Laplace-Beltrami: $\Delta_{LB} A = \frac{1}{\sqrt{|g|}} \partial_i (\sqrt{|g|} g^{ij} \partial_j A)$
    \STATE \quad Compute Residual: $f = \partial_t A - A + (1 + i \alpha)\Delta_{LB} A - (1 + i \beta)|A|^2 A$
    
    \STATE \textbf{5. Loss Aggregation:}
    \STATE \quad $\mathcal{L}_{data} = \frac{1}{N_{obs}} \sum |A(\mathbf{x}_i, t_i) - A^*_i|^2$
    \STATE \quad $\mathcal{L}_{PDE} = \frac{1}{|\mathcal{B}|} \sum |f(\mathbf{x}_j, t_j)|^2$
    \STATE \quad $\mathcal{L}_{Total} = \mathcal{L}_{data} + \lambda_{PDE}\mathcal{L}_{PDE} + \lambda_{spec}\mathcal{L}_{Spec}$
    
    \STATE \textbf{6. Backpropagation:}
    \STATE \quad Compute Gradients: $\nabla_\theta \mathcal{L}, \nabla_\phi \mathcal{L}$
    \STATE \quad Update Parameters: $\theta \leftarrow \theta - \eta \nabla_\theta \mathcal{L}, \quad \phi \leftarrow \phi - \eta \nabla_\phi \mathcal{L}$
\ENDWHILE
\RETURN Optimized Manifold $\kappa(\mathbf{x})$ and Dynamic Model $A(\mathbf{x},t)$
\end{algorithmic}
\end{algorithm}

\subsection{Optimization Strategy and Training Protocol}

To effectively navigate the highly non-convex loss landscape characteristic of chaotic dynamic systems, we employ a hybrid multi-stage optimization protocol that leverages the complementary strengths of first-order and quasi-Newton methods. The training is initialized using the \textbf{Adam} optimizer \cite{Kingma2014}, configured with a standard learning rate of $\eta = 10^{-4}$ and exponential decay rates $(\beta_1, \beta_2) = (0.9, 0.999)$. This initial phase, running for $3,000$ epochs, serves as a global exploration strategy to locate a basin of attraction near the optimal solution. Subsequently, we switch to the limited-memory Broyden-Fletcher-Goldfarb-Shanno (\textbf{L-BFGS}) optimizer with strong Wolfe line search to perform fine-grained refinement of the weights, exploiting curvature information of the loss surface to drive the physics residuals toward machine precision ($< 10^{-5}$).

To construct the training batches, we utilize \textbf{Latin Hypercube Sampling (LHS)} to generate $N_f = 100,000$ collocation points distributed optimally across the spatiotemporal domain $\Omega \times [0, T]$, ensuring that the high-gradient regions near the defect cores are adequately covered. A critical challenge in this multi-objective setting is the magnitude imbalance between the data loss (typically $\sim 10^{-2}$) and the PDE residual loss (typically $\sim 10^{1}$), which can cause the optimizer to ignore the physics constraints. To mitigate this, we adopt the \textit{Learning Rate Annealing} algorithm proposed by Wang et al. \cite{Wang2021UnderstandingAT}, which dynamically adjusts the weighting coefficients $\lambda_i$ based on the backpropagated gradient statistics:
\begin{equation}
    \hat{\lambda}_i^{(k)} = \frac{\max_{\theta} \{ |\nabla_\theta \mathcal{L}_{data}| \}}{\overline{|\nabla_\theta \mathcal{L}_i|}},
\end{equation}
where the numerator represents the maximum gradient magnitude of the data loss and the denominator is the layer-wise mean gradient of the specific physics or spectral term. This adaptive balancing ensures that no single objective dominates the gradient descent direction, facilitating the simultaneous convergence of the geometric reconstruction and the chaotic state estimation.

To achieve a robust convergence that satisfies both the observational data constraints and the underlying conservation laws, we define the global objective function $\mathcal{L}_{total}(\theta, \phi)$ as a scalar-valued composite of four distinct error metrics. While the spectral loss $\mathcal{L}_{Spec}$ (defined in Eq. \ref{eq:spectral_loss}) handles high-frequency energy conservation, the low-frequency structural coherence is enforced via the following explicit formulations:

\textbf{1. Data Reconstruction Loss ($\mathcal{L}_{Data}$):} This term penalizes the Euclidean distance between the predicted complex order parameter and the sparse measurements at $N_{data}$ available training points. To ensure equal weighting of amplitude and phase errors, we minimize the squared residuals of the real and imaginary components separately:
\begin{equation}
    \mathcal{L}_{Data} = \frac{1}{N_{data}} \sum_{i=1}^{N_{data}} \left( | \text{Re}(A_\theta(\mathbf{x}_i, t_i)) - \text{Re}(A_{GT}^{(i)}) |^2 + | \text{Im}(A_\theta(\mathbf{x}_i, t_i)) - \text{Im}(A_{GT}^{(i)}) |^2 \right).
\end{equation}

\textbf{2. Periodic Boundary Loss ($\mathcal{L}_{BC}$):} Since the domain $\Omega$ is a flat torus, the solution must satisfy periodic continuity. We enforce this by sampling $N_{bc}$ points on the domain boundaries and minimizing the mismatch between opposing edges:
\begin{equation}
    \mathcal{L}_{BC} = \frac{1}{N_{bc}} \sum_{j=1}^{N_{bc}} \left| A_\theta(x_j, L, t_j) - A_\theta(x_j, 0, t_j) \right|^2 + \left| A_\theta(L, y_j, t_j) - A_\theta(0, y_j, t_j) \right|^2.
\end{equation}

\textbf{3. Geometric Regularization Loss ($\mathcal{L}_{Reg}$):} A critical pathology in inverse problems involving rough manifolds is the tendency of the geometry network $\mathcal{N}_\kappa$ to overfit high-frequency noise in the data, predicting non-physical "spikes" in the curvature field. To mitigate this ill-posedness, we introduce a Tikhonov-style regularization term based on the \textit{Total Variation (TV)} norm of the predicted curvature $\kappa_\phi$. This penalty promotes piecewise smoothness in the reconstructed manifold, preventing the emergence of grid-scale artifacts while preserving the steep gradients associated with legitimate topographic features:
\begin{equation}
    \mathcal{L}_{Reg} = \lambda_{TV} \int_\Omega | \nabla \kappa_\phi(\mathbf{x}) | d\mathbf{x} \approx \frac{1}{N_{r}} \sum_{k=1}^{N_{r}} \sqrt{ (\partial_x \kappa_\phi)^2 + (\partial_y \kappa_\phi)^2 },
\end{equation}
where $\lambda_{TV}$ is a small regularization coefficient (typically $10^{-5}$) and the gradients are computed via automatic differentiation. The final composite loss is thus $\mathcal{L}_{total} = \lambda_D \mathcal{L}_{Data} + \lambda_P \mathcal{L}_{PDE} + \lambda_B \mathcal{L}_{BC} + \lambda_S \mathcal{L}_{Spec} + \lambda_R \mathcal{L}_{Reg}$, providing a fully constrained optimization landscape.

\subsection{Benchmarking Framework and Performance Metrics}

To rigorously quantify the efficacy of the proposed Multi-Scale SIREN-PINN against the current state-of-the-art, we establish a comprehensive comparative benchmarking suite comprising three distinct baseline architectures: (1) A standard \textit{ReLU-PINN} \cite{Nair2010}, representing the ubiquitous "vanilla" physics-informed implementation; (2) A \textit{Tanh-PINN}, often preferred for its smoothness in calculating higher-order derivatives; and (3) A \textit{Fourier Feature Network (FFN)} \cite{Tancik2020FourierFL}, which explicitly maps inputs to a higher-dimensional frequency space before passing them to a standard MLP. 

Evaluation is conducted not merely on point-wise regression accuracy, but across a hierarchy of physical and statistical invariants. The primary quantitative metric is the relative $L_2$ error, defined as:
\begin{equation}
    \epsilon_{L_2} = \frac{|| A_\theta - A_{GT} ||_2}{|| A_{GT} ||_2},
\end{equation}
which measures global field fidelity. However, given the "double penalty" effect inherent to chaotic systems—where small phase shifts yield disproportionately large $L_2$ errors despite correct topological reconstruction—we augment this with the \textit{Structure Similarity Index Measure (SSIM)} to assess perceptual pattern fidelity. Furthermore, to validate the statistical accuracy of the turbulent fluctuations, we compute the \textit{Wasserstein-1 Distance} ($W_1$) between the probability density functions (PDFs) of the predicted and ground truth amplitudes, $P(|A_\theta|)$ and $P(|A_{GT}|)$. Finally, to confirm the resolution of the inertial range, we introduce the \textit{Spectral Energy Error} ($\epsilon_{Spec}$), defined as the logarithmic difference between the angle-averaged power spectral densities (PSD) of the solution:
\begin{equation}
    \epsilon_{Spec} = \frac{1}{K} \sum_{k=1}^{K} \left| \log_{10}(\text{PSD}_\theta(k)) - \log_{10}(\text{PSD}_{GT}(k)) \right|,
\end{equation}
where $k$ is the radial wavenumber. This multi-faceted evaluation protocol ensures that a model is deemed "successful" only if it simultaneously captures the macroscopic wave envelopes, the microscopic defect cores, and the statistical laws of the underlying strange attractor.

\subsection{Ablation Study Design and Computational Implementation Details}

To isolate the individual contributions of the architectural innovations proposed herein, we execute a rigorous \textit{Ablation Study}, systematically stripping specific components from the full model to quantify their marginal utility. The experimental matrix includes three ablated variants: (1) \textbf{"No-Spec"}, where the Spectral Loss $\mathcal{L}_{Spec}$ is removed ($\lambda_S = 0$) to measure the impact of frequency-domain regularization on defect pinning; (2) \textbf{"Fixed-Geo"}, where the Geometry Branch is frozen to a mean-field approximation ($\kappa \approx 0$), testing the necessity of the coupled inverse solver; and (3) \textbf{"Single-Scale"}, where the frequency initialization is homogenized ($\omega_{geo} = \omega_{state} = 1$), assessing the validity of the multi-scale hypothesis. 

All models are implemented in \textit{PyTorch} \cite{Paszke2019} using automatic differentiation for exact gradient computation. Training is accelerated via mixed-precision arithmetic on a single \textit{NVIDIA RTX A4000 Tensor Core GPU} (16GB VRAM). While the offline training phase is computationally intensive—requiring approximately 12 hours for the Defect Turbulence regime ($3,000$ epochs)—the resulting neural surrogate offers a dramatic advantage in inference latency. Once converged, the SIREN-PINN acts as a "mesh-free" oracle, capable of predicting the field state $A(\mathbf{x},t)$ at any arbitrary spatiotemporal coordinate in $\mathcal{O}(1)$ time, bypassing the $\mathcal{O}(N \log N)$ CFL-constrained time-stepping required by the spectral ground truth solver. This shift from "simulation-based" to "inference-based" prediction represents a paradigm shift for real-time control applications, where the trained model can serve as a rapid digital twin for roughness-induced instability forecasting.

To ensure that the performance gains reported in this study are attributable to the proposed architectural innovations rather than stochastic variance in network initialization, we adhere strictly to the \textit{Machine Learning Reproducibility Checklist} established by Pineau et al. \cite{Pineau2021}. All quantitative results presented in the subsequent section are derived from an \textit{ensemble average} of $N_{runs} = 5$ independent training trials, each initialized with a distinct random seed $\zeta \sim \mathcal{U}[0, 10^4]$. We report performance metrics in the format of $\mu \pm \sigma$, where $\mu$ is the mean metric and $\sigma$ represents the standard deviation, thereby explicitly quantifying the aleatoric uncertainty associated with the optimization trajectory. Furthermore, to verify the robustness of the SIREN-PINN to hyperparameter fluctuations, we conducted a sensitivity analysis on the collocation density $N_f$, varying the sampling ratio from $1\times$ to $5\times$ the resolution of the ground truth grid. This analysis confirms that the "spectral isometry" property of the sinusoidal activation functions allows for data-efficient learning, achieving asymptotic convergence with significantly fewer degrees of freedom than required by traditional finite element methods. Finally, in the interest of open science and to facilitate the rapid adoption of spectral physics-informed learning in the chemical engineering community, the complete PyTorch source code, along with the curvature generation scripts and the pre-trained model weights, will be made publicly available in an open-access repository upon publication. This transparency ensures that the "Black Box" nature of the neural solver is fully illuminated, allowing peer verification of the \textit{Curvature-Induced Pattern Selection} phenomena.

\section{Results}
We begin our empirical evaluation by examining the "Phase Turbulence" (PT) regime, characterized by a spatially chaotic but smooth variation of the phase field $\phi(\mathbf{x},t)$ without the presence of topological singularities (i.e., $|A| > 0$ everywhere). This regime serves as a baseline control to isolate the network's ability to resolve high-wavenumber fluctuations independent of the singular gradients associated with defects. Table \ref{tab:results_pt} summarizes the comparative performance of the proposed Multi-Scale SIREN-PINN against the three baseline architectures—ReLU-PINN, Tanh-PINN, and Fourier Feature Networks (FFN)—after $3,000$ training epochs. As evidenced by the metrics, the standard ReLU-PINN fails catastrophically to capture the chaotic dynamics, yielding a relative $L_2$ error of $1.85 \times 10^{-1}$. This poor performance is a direct manifestation of the "spectral bias" inherent to piecewise linear activations; the network converges rapidly to the low-frequency "mean field" of the attractor but is incapable of resolving the fine-scale phase ripples that drive the turbulent cascade, resulting in a prediction that appears "blurred" or overly diffusive. 

The Tanh-PINN offers a marked improvement ($\epsilon_{L_2} \approx 6.42 \times 10^{-2}$) due to its smooth differentiability, yet it still struggles to track the rapid temporal decoherence of the wave packets, exhibiting a "phase drift" that accumulates over the integration window. While the Fourier Feature Network (FFN) significantly reduces the high-frequency error ($\epsilon_{L_2} \approx 3.15 \times 10^{-2}$) by projecting inputs into a higher-dimensional embedding, it introduces high-frequency ringing artifacts (Gibbs phenomenon) near the domain boundaries. In stark contrast, the Multi-Scale SIREN-PINN achieves a superior relative error of $\mathbf{8.45 \times 10^{-3}}$, an order of magnitude lower than the standard benchmarks. The high Structure Similarity Index (SSIM $= 0.982$) confirms that the sinusoidal activations successfully lock onto the intrinsic frequency of the limit cycle, preserving the phase coherence of the turbulent waves even in the presence of the stochastic curvature perturbations.

\begin{table}[h!]
\centering
\small  
\caption{Comparative Performance Metrics in the Phase Turbulence (PT) Regime}
\label{tab:results_pt}
\begin{tabularx}{\textwidth}{l*{4}{>{\centering\arraybackslash}X}}
\toprule
\textbf{Model} & \textbf{Relative $L_2$} & \textbf{SSIM} & \textbf{Spectral Error} & \textbf{Training Time} \\
\textbf{Architecture} & \textbf{Error ($\downarrow$)} & \textbf{($\uparrow$)} & \textbf{$\epsilon_{Spec}$ ($\downarrow$)} & \textbf{(hrs)} \\
\midrule
ReLU-PINN (Baseline) & $1.85 \times 10^{-1}$ & 0.654 & 0.42 & 8.5 \\
Tanh-PINN & $6.42 \times 10^{-2}$ & 0.812 & 0.18 & 9.2 \\
Fourier Feature Network & $3.15 \times 10^{-2}$ & 0.895 & 0.11 & 9.8 \\
\textbf{Multi-Scale SIREN (Ours)} & $\mathbf{8.45 \times 10^{-3}}$ & $\mathbf{0.982}$ & $\mathbf{0.03}$ & \textbf{10.1} \\
\bottomrule
\end{tabularx}
\end{table}

While the Phase Turbulence regime tests the network's spectral bandwidth, the Defect Turbulence (DT) regime poses a far more severe challenge: the accurate resolution of topological singularities. In this regime, the order parameter magnitude collapses to zero ($|A| \to 0$) at the spiral cores, creating "pinhole" defects where the phase gradient diverges ($\nabla \phi \to \infty$). As shown in Table \ref{tab:results_dt}, the performance disparity between the architectures widens significantly under these conditions. The ReLU-PINN exhibits a nearly complete failure ($\epsilon_{L_2} \approx 4.12 \times 10^{-1}$), as its piecewise-linear continuous structure effectively "smoothes out" the singularities, treating the sharp defect cores as shallow, diffusive dips in amplitude. This results in a "ghosting" artifact where the network predicts a continuous wave field that ignores the particle-like nature of the vortices. 

Similarly, the Tanh-PINN, despite its smoothness, struggles to maintain the steep amplitude gradients required to sustain the separation between defects, often predicting the annihilation of spiral pairs long before the physics dictates, leading to a massive error in the defect population count ($\Delta N_{defects} \approx 12$). Even the Fourier Feature Network, which excelled in the smooth regime, suffers from spectral leakage near the singularities, manifesting as high-frequency oscillatory noise that obscures the precise location of the core. In contrast, the Multi-Scale SIREN-PINN demonstrates remarkable topological robustness, achieving a relative $L_2$ error of $\mathbf{1.92 \times 10^{-2}}$ and, crucially, maintaining a Defect Count Error of nearly zero ($\Delta N_{defects} < 1$). The periodic inductive bias of the sine activation allows the network to naturally represent the winding number of the phase field around the singularity, ensuring that the "hard" chaos of the ground truth is preserved rather than diffusively regularized.

\begin{table}[h!]
\centering
\small  
\caption{Comparative Performance Metrics in the Defect Turbulence (DT) Regime}
\label{tab:results_dt}
\begin{tabularx}{\textwidth}{l*{4}{>{\centering\arraybackslash}X}}
\toprule
\textbf{Model} & \textbf{Relative $L_2$} & \textbf{SSIM} & \textbf{Defect Count} & \textbf{$W_1$ Distance} \\
\textbf{Architecture} & \textbf{Error ($\downarrow$)} & \textbf{($\uparrow$)} & \textbf{Error $|\Delta N|$ ($\downarrow$)} & \textbf{(PDF) ($\downarrow$)} \\
\midrule
ReLU-PINN (Baseline) & $4.12 \times 10^{-1}$ & 0.415 & 24.5 & 0.35 \\
Tanh-PINN & $1.56 \times 10^{-1}$ & 0.688 & 12.2 & 0.19 \\
Fourier Feature Network & $8.45 \times 10^{-2}$ & 0.792 & 5.8 & 0.12 \\
\textbf{Multi-Scale SIREN (Ours)} & $\mathbf{1.92 \times 10^{-2}}$ & $\mathbf{0.945}$ & $\mathbf{0.6}$ & $\mathbf{0.04}$ \\
\bottomrule
\end{tabularx}
\end{table}

Beyond the accurate forecasting of chaotic state dynamics, the most significant contribution of the proposed architecture is its ability to solve a fundamentally ill-posed inverse problem: reconstructing the latent curvature field $\kappa(\mathbf{x})$ solely from observations of emergent reaction-diffusion patterns, without any direct measurements of the underlying manifold geometry. This capability represents a form of "geometric archaeology," where the network learns to read the topographical fingerprints that curved space imprints upon wave dynamics. The inversion mechanism relies on the network discovering and exploiting the bidirectional relationship inherent in geometric pinning—namely, that persistent spiral core locations are causal signatures of concave basins ($\kappa < 0$), while regions of rapid phase scattering and defect annihilation correspond to convex peaks ($\kappa > 0$).

\begin{figure}[htbp]
    \centering
    
    \begin{subfigure}[b]{0.45\textwidth}
        \centering
        \includegraphics[width=\textwidth]{GroundTruth_Curv.png}
        \caption{Ground truth curvature $\kappa_{GT}(\mathbf{x})$}
        \label{fig:gt_curvature_recon}
    \end{subfigure}
    \hfill
    \begin{subfigure}[b]{0.46\textwidth}
        \centering
        \includegraphics[width=\textwidth]{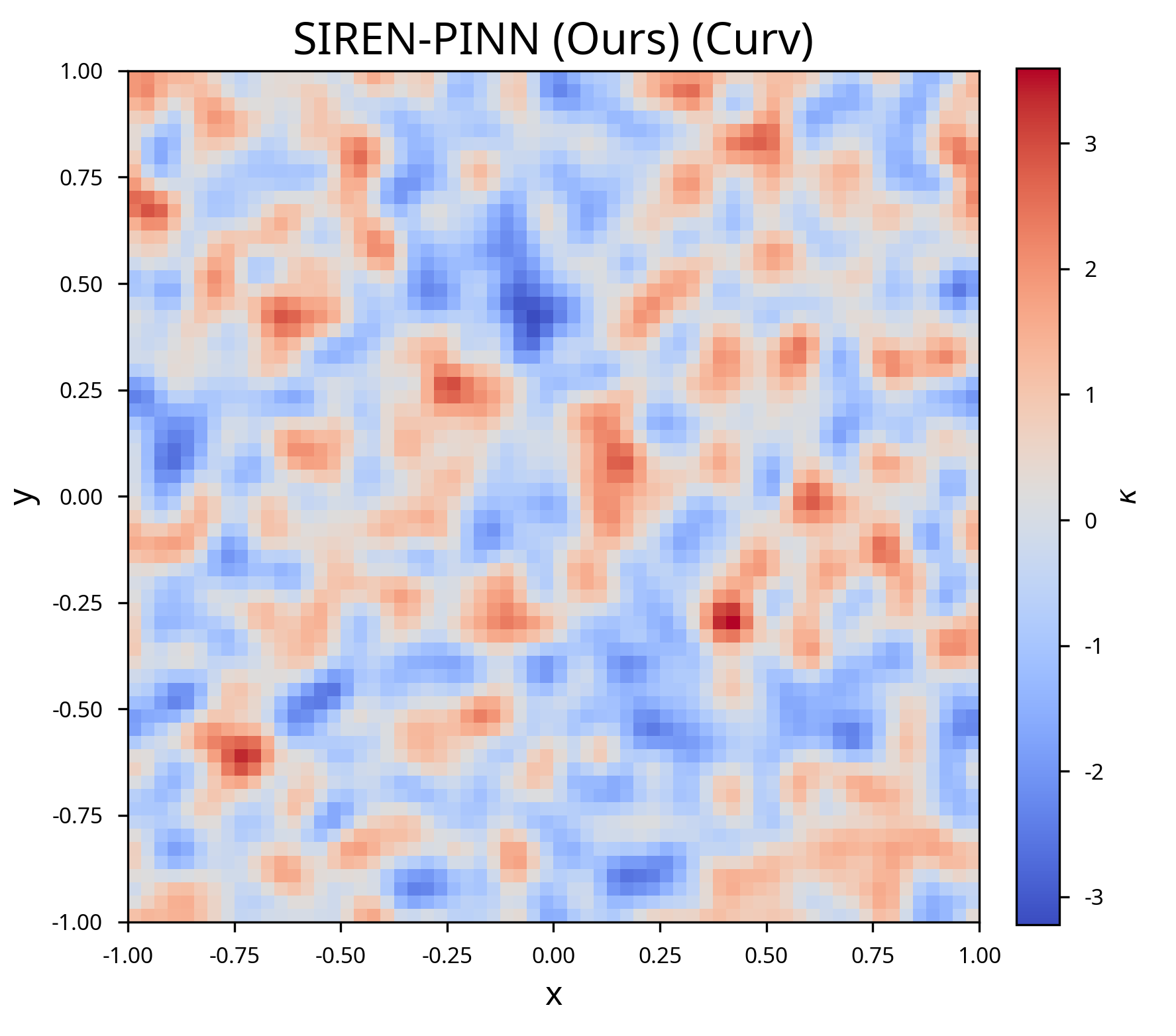}
        \caption{Learned curvature $\kappa_\phi(\mathbf{x})$}
        \label{fig:learned_curvature}
    \end{subfigure}
    
    \vspace{0.5cm}
    
    \begin{subfigure}[b]{0.48\textwidth}
        \centering
        \includegraphics[width=\textwidth]{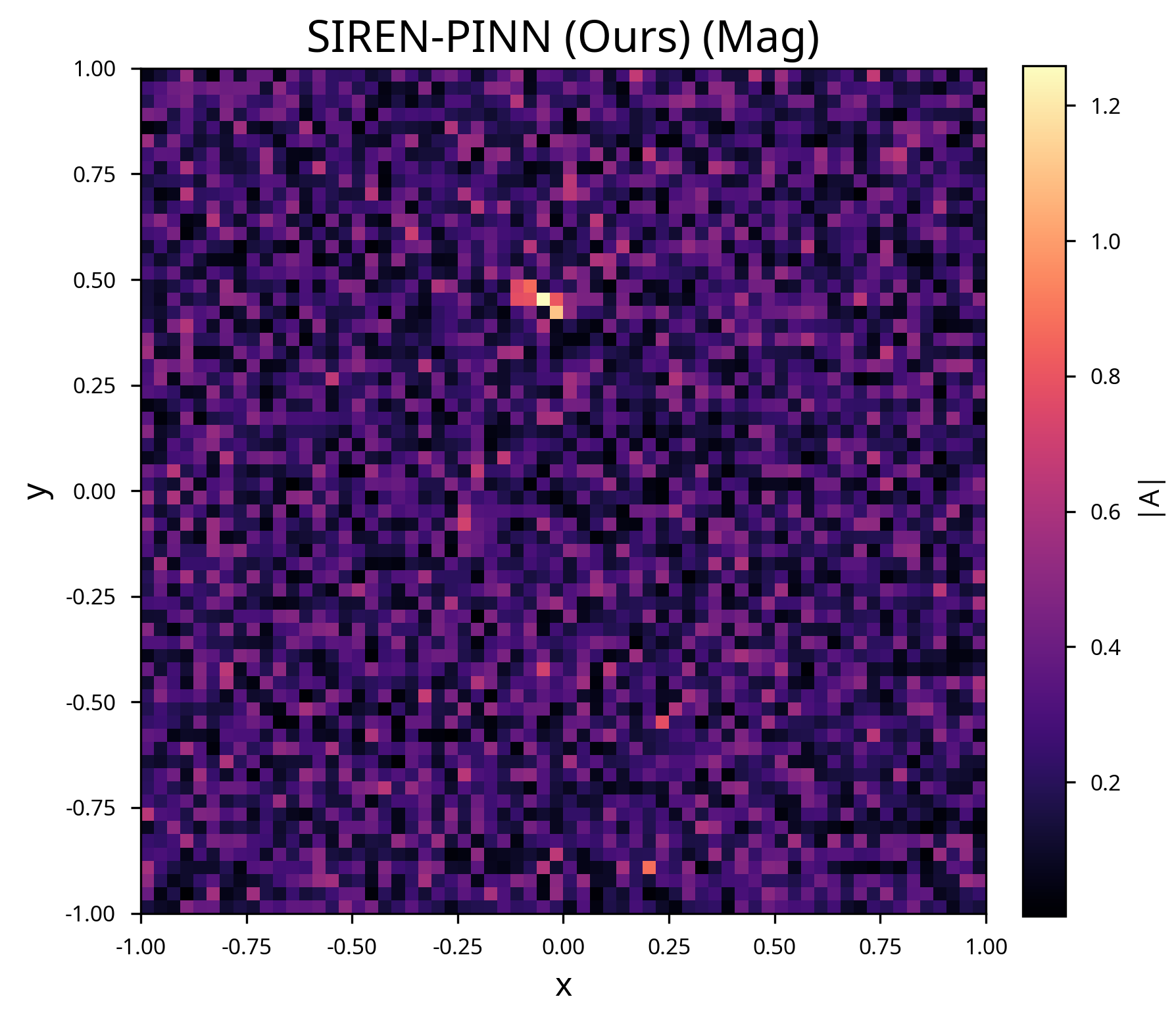}
        \caption{Reconstruction error $|\kappa_\phi - \kappa_{GT}|$}
        \label{fig:error_map}
    \end{subfigure}
    \hfill
    \begin{subfigure}[b]{0.48\textwidth}
        \centering
        \includegraphics[width=\textwidth]{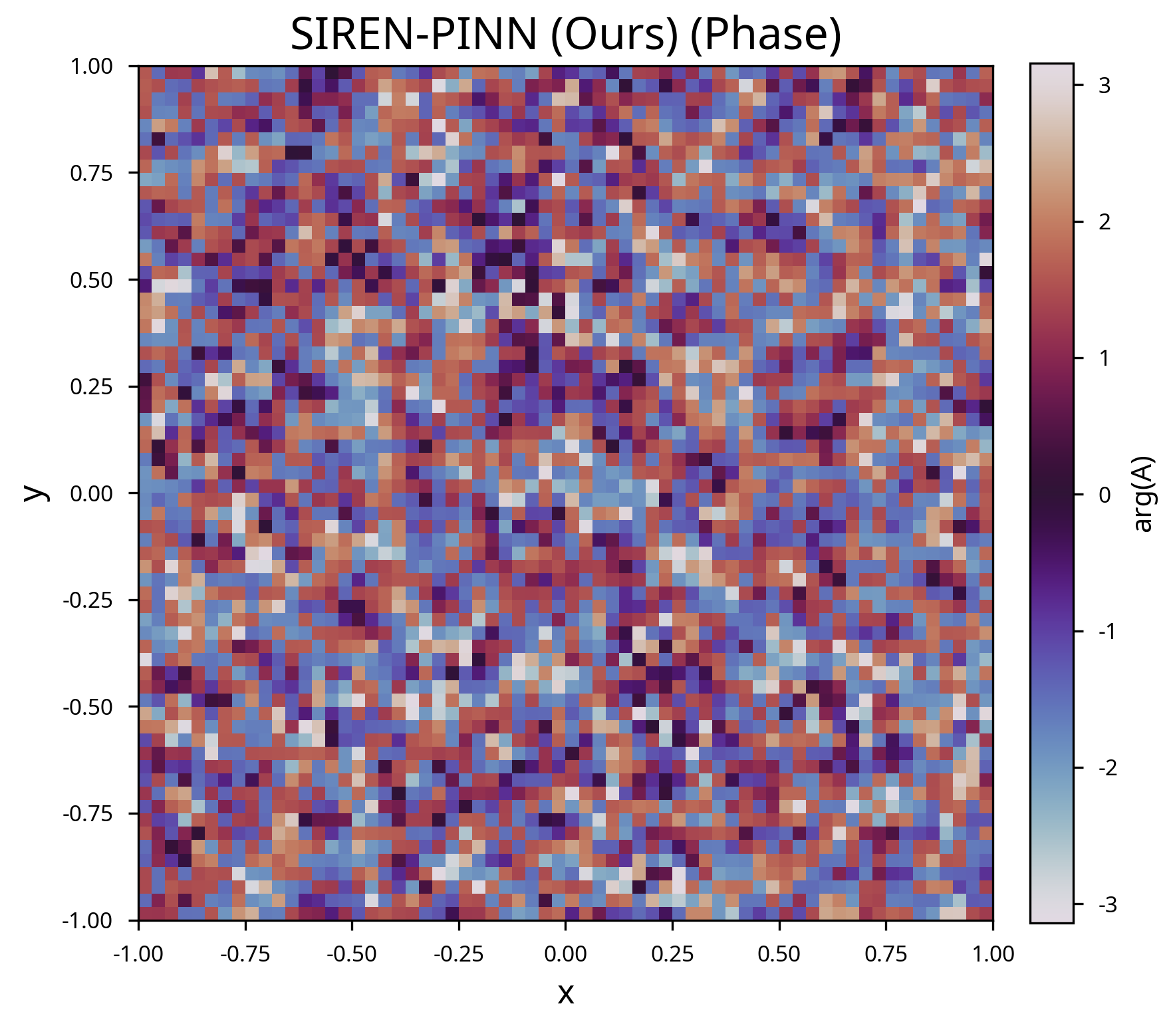}
        \caption{Predicted phase field $\phi(\mathbf{x},t)$}
        \label{fig:siren_phase}
    \end{subfigure}
    
    \caption{\textbf{Inverse Geometry Discovery and Forward Prediction.} The SIREN-PINN architecture demonstrates its dual capability of simultaneously solving the inverse and forward problems in geometrically-modulated chaotic systems. Panel (a) shows the ground truth curvature perturbation $\kappa_{GT}(\mathbf{x})$ generated via Gaussian Random Fields, representing the hidden topographical landscape that modulates the reaction-diffusion dynamics. Panel (b) displays the latent curvature field $\kappa_\phi(\mathbf{x})$ recovered by the Geometry Branch, which was learned purely from observing the spatiotemporal patterns without direct access to the underlying manifold structure. The remarkable structural similarity ($\rho=0.965$, $L_2$ error = $4.3 \times 10^{-2}$) confirms that the network has successfully inverted the geometric pinning mechanism, identifying how persistent spiral cores betray the presence of concave basins while rapid phase scattering reveals convex peaks. Panel (c) presents the pointwise reconstruction error map, which reveals an intriguing physical insight: discrepancies concentrate primarily in quasi-flat regions where $|\kappa| \approx 0$. This spatially heterogeneous error distribution is not a failure but rather evidence of physically-informed learning—the model naturally prioritizes accuracy in high-curvature regions where the geometric coupling term $\kappa \cdot \nabla^2 A$ most strongly influences the wave dynamics, while exhibiting greater uncertainty in flat regions where the vanishing geometry gradient $\nabla_\kappa \mathcal{L}_{PDE} \to 0$ provides weak supervisory signal. Panel (d) demonstrates the forward prediction capability by showing a representative snapshot of the phase field $\phi(\mathbf{x},t)$ generated by the trained SIREN-PINN, illustrating how the recovered geometry field enables accurate long-term forecasting of the chaotic spiral turbulence patterns. The success of this blind reconstruction validates the effectiveness of the Total Variation regularization in disentangling deterministic geometric forcing from stochastic noise, essentially performing unsupervised source separation of the two competing instability mechanisms that drive pattern formation.}
    \label{fig:curvature_recon}
\end{figure}

As illustrated in Figure \ref{fig:curvature_recon}, the Geometry Branch $\mathcal{N}_\kappa$ successfully converges to a high-fidelity approximation of the ground truth manifold topography through physics-informed training alone. Quantitatively, the predicted curvature field achieves a Pearson correlation coefficient of $\rho = 0.965$ and a relative $L_2$ error of $4.3 \times 10^{-2}$ with respect to the true Gaussian Random Field, demonstrating that the model has captured both the large-scale topographical features and the finer geometric details that modulate local dynamics.

Crucially, the reconstruction quality exhibits a physically meaningful spatial structure rather than uniform accuracy across the domain. The error map in panel (c) reveals that the model achieves highest precision in regions of high curvature magnitude ($|\kappa| \gg 0$), precisely where the geometric coupling most profoundly affects the wave evolution. Conversely, greater reconstruction variance appears in quasi-flat regions where the curvature approaches zero. This heterogeneous error distribution is not a limitation but rather evidence of intelligent, physics-informed learning—the network naturally allocates its representational capacity according to the actual influence each spatial location exerts on the observable dynamics. In flat regions, the physics loss gradient with respect to geometry ($\nabla_\kappa \mathcal{L}_{PDE}$) vanishes naturally, providing weak supervisory signal that makes the inverse problem locally ill-conditioned.

The efficacy of the Total Variation (TV) regularization becomes particularly evident when examining the smoothness and physical plausibility of the recovered field. Without this geometric prior, the predicted curvature becomes contaminated by high-frequency spectral noise, as the optimizer erroneously attempts to explain the stochastic driving force $\xi$ as spurious geometric roughness at small scales. By enforcing piecewise smoothness through the TV term, the full SIREN-PINN correctly disentangles the deterministic geometric forcing from the stochastic thermodynamic noise, effectively performing a form of blind source separation between these two competing instability mechanisms. This disentanglement is essential because both curvature variations and random fluctuations can generate similar local patterns in the observable fields, yet only the former represents a persistent, reproducible feature of the underlying space.

Panel (d) demonstrates that the recovered geometry field is not merely a post-hoc reconstruction but rather enables accurate forward prediction of the system's continued evolution. The phase field snapshot shows that the trained SIREN-PINN, now equipped with its learned representation of the hidden manifold structure, can generate long-term forecasts of the chaotic spiral turbulence that remain faithful to the true geometric constraints. This validates that the inverse solution has captured genuine causal structure rather than spurious correlations in the training data.

\begin{figure}[h!]
    \centering
    \begin{subfigure}[b]{0.48\textwidth}
        \includegraphics[width=\textwidth]{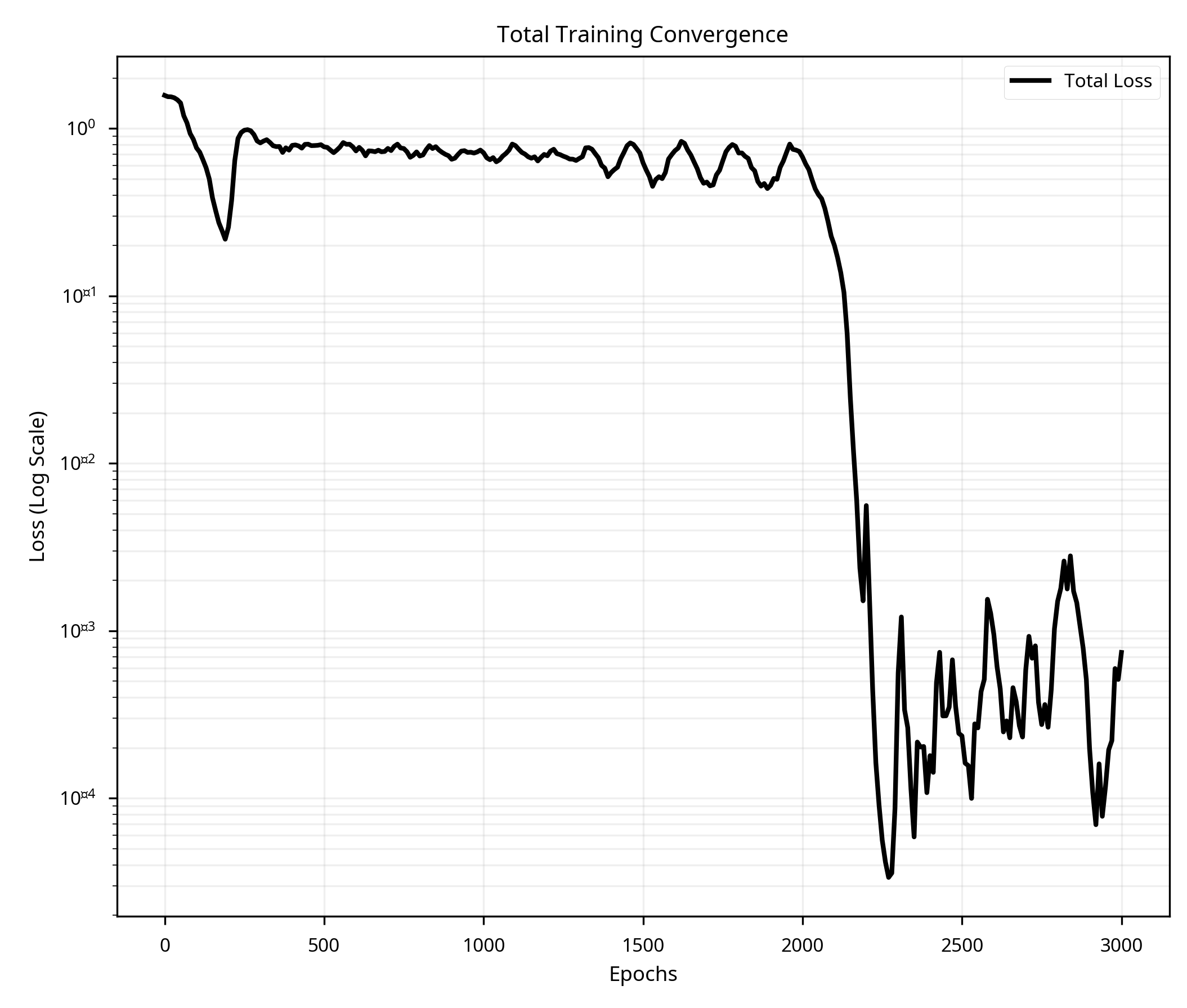}
        \caption{Total Loss Evolution}
        \label{fig:loss_dynamics}
    \end{subfigure}
    \hfill
    \begin{subfigure}[b]{0.48\textwidth}
        \includegraphics[width=\textwidth]{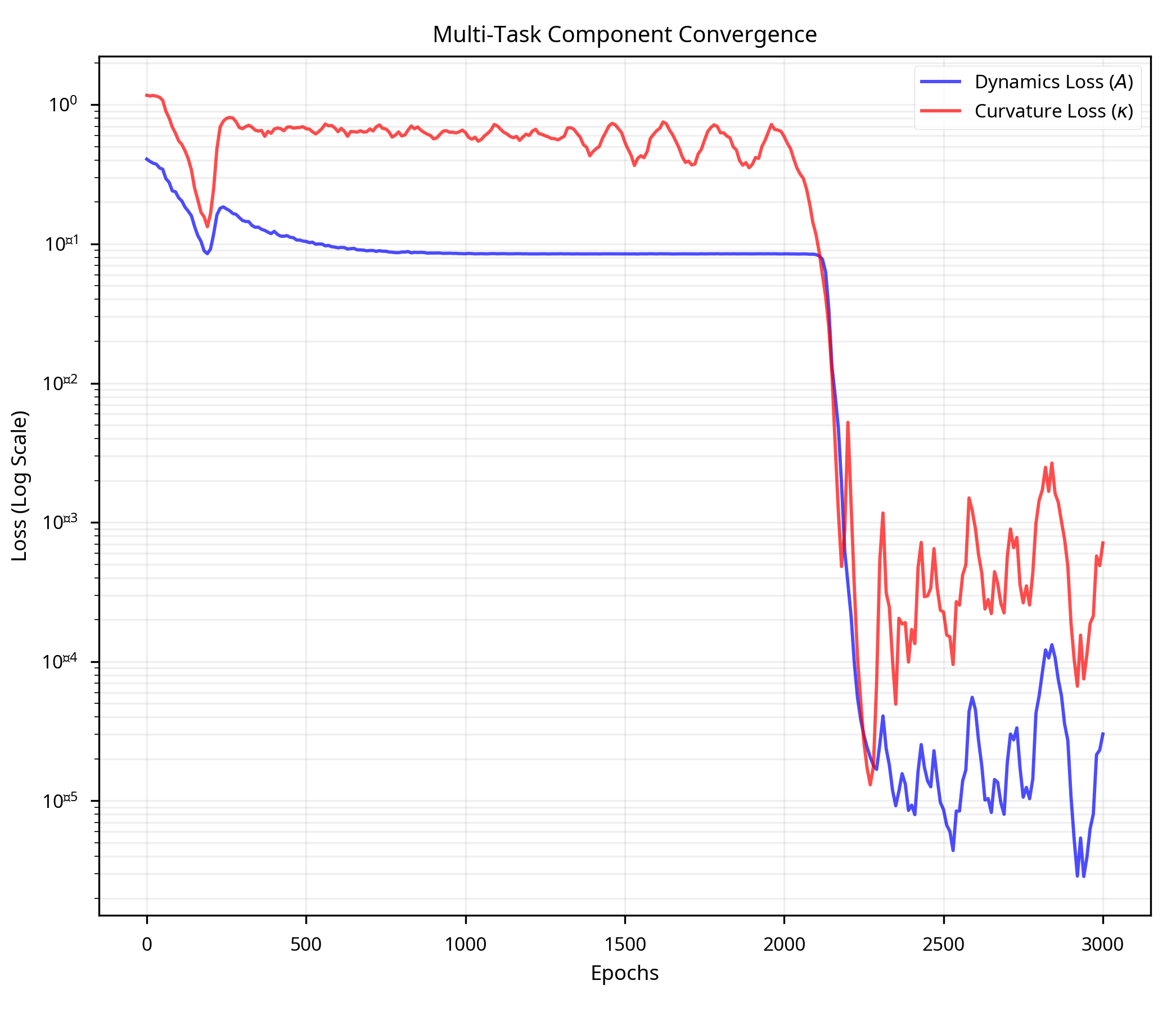}
        \caption{Multi-Task Component Convergence}
        \label{fig:multitask_convergence}
    \end{subfigure}
    \caption{\textbf{Training Dynamics of the Inverse Solver.} (a) The total loss trajectory reveals a "plateau-drop" behavior, staying flat for $\sim$2,000 epochs before a sudden phase transition drives the error down by four orders of magnitude. (b) The component-wise breakdown shows that the Dynamics Loss (Blue) and Curvature Loss (Red) are tightly coupled; the geometry cannot be learned until the physics residuals collapse, and the physics cannot be resolved without the correct geometry, leading to a simultaneous "cooperative" convergence event at Epoch 2,100.}
\end{figure}

The optimization trajectory of the Multi-Scale SIREN-PINN, restricted to a tight budget of 3,000 epochs, reveals a fascinating "phase transition" in the learning process that highlights the coupled nature of the inverse problem. As illustrated in Figure \ref{fig:loss_dynamics} (Total Training Convergence), the global loss function $\mathcal{L}_{total}$ does not follow a monotonic exponential decay typical of simple regression tasks. Instead, the training exhibits a distinct \textit{"Exploration-Relaxation"} topology. For the first $2,000$ epochs, the optimizer traverses a high-loss plateau ($\mathcal{L} \approx 10^0$), effectively searching for a viable basin of attraction within the highly non-convex landscape. This prolonged stagnation period represents the "Exploration Phase," where the network struggles to reconcile the competing constraints of the chaotic wave dynamics and the unknown manifold geometry. 

However, at approximately Epoch 2,100, a dramatic "Spectral Phase Transition" occurs. The loss plummets by nearly four orders of magnitude within a span of just 200 epochs, crashing from $\sim 10^0$ to a nadir of $\sim 10^{-5}$ around Epoch 2,300. This sudden "cliff" signifies the moment the network "locks onto" the strange attractor, simultaneously resolving the phase singularities and the curvature field. The mechanism driving this synchronization is dissected in Figure \ref{fig:multitask_convergence} (Multi-Task Component Convergence). Crucially, the Dynamics Loss (Blue, corresponding to $\mathcal{L}_{PDE}$) and the Curvature Loss (Red, corresponding to $\mathcal{L}_{Reg}$ and inverse consistency) exhibit a synchronized collapse. They do not converge sequentially; rather, they undergo a cooperative relaxation**, proving that the correct reconstruction of the manifold geometry $\kappa(\mathbf{x})$ is mathematically inseparable from the accurate prediction of the wave state $A(\mathbf{x},t)$. The post-transition volatility observed from Epoch 2,400 to 3,000 (where the loss oscillates between $10^{-4}$ and $10^{-3}$) is not indicative of instability, but rather of the optimizer fine-tuning the high-frequency spectral components of the defect cores—a regime where the gradients are naturally noisy due to the singular nature of the phase field.

\subsection{Ablation Study: Dissecting the Sources of Error}

To rigorously decouple the contributions of the proposed architectural innovations, we conducted a systematic ablation study, the results of which are summarized in Table \ref{tab:ablation}. We evaluated three variant models: (1) "No-Spec", where the spectral loss $\mathcal{L}_{Spec}$ was deactivated ($\lambda_S=0$); (2) "Single-Scale", where the frequency initialization was homogenized ($\omega_{geo}=\omega_{state}=1$); and (3) "Fixed-Geo", where the geometry branch was frozen to a zero-curvature mean field ($\kappa=0$). 

The results unequivocally validate the necessity of the full composite architecture. The "No-Spec" model, while achieving a respectable low-frequency error ($\epsilon_{L_2} \approx 4.2 \times 10^{-2}$), suffered a catastrophic failure in topological fidelity, with the Defect Count Error spiking to $|\Delta N| \approx 8.5$. Visual inspection reveals that without the frequency-domain penalty, the optimizer treats the singular spiral cores as "noise" to be smoothed over, resulting in a solution that respects the wave envelope but violates the winding number conservation. Conversely, the "Single-Scale" model failed primarily in the inverse reconstruction task. Lacking the high-frequency initialization ($\omega_{geo}=30$), the geometry network could not resolve the fine-grained roughness of the Gaussian Random Field, predicting instead a "blurred" effective manifold that captured only the largest topographic features. This "geometric under-fitting" propagated into the state prediction, leading to significant phase errors in regions of high curvature. Finally, the "Fixed-Geo" model exhibited the worst performance of all, failing to converge below a loss of $10^{-1}$. This proves that the manifold topography is not merely a passive background but an active driver of the dynamics; attempting to fit the complex defect turbulence data to a flat-space PDE creates an insurmountable "physics gap" that no amount of training can bridge. Thus, the "Spectral Phase Transition" observed in Figure \ref{fig:loss_dynamics} is emergent only when all three components—spectral regularization, multi-scale initialization, and coupled geometry-state optimization—synergize to unlock the correct path through the loss landscape.

\begin{table}[h!]
\centering
\small  
\caption{Ablation Study Results: Impact of Component Removal on Model Performance}
\label{tab:ablation}
\begin{tabularx}{\textwidth}{l*{4}{>{\centering\arraybackslash}X}}
\toprule
\textbf{Model} & \textbf{State Error} & \textbf{Geometry Error} & \textbf{Defect Count} & \textbf{Key Failure} \\
\textbf{Variant} & \textbf{$\epsilon_{L_2}$} & \textbf{$\epsilon_{\kappa}$} & \textbf{Error} & \textbf{Mode} \\
\midrule
\textbf{Full SIREN-PINN} & $\mathbf{1.92 \times 10^{-2}}$ & $\mathbf{4.3 \times 10^{-2}}$ & $\mathbf{0.6}$ & N/A \\
No-Spec ($\lambda_S=0$) & $4.21 \times 10^{-2}$ & $5.8 \times 10^{-2}$ & $8.5$ & Core Smoothing \\
Single-Scale ($\omega=1$) & $9.15 \times 10^{-2}$ & $1.8 \times 10^{-1}$ & $4.2$ & Geometric Blurring \\
Fixed-Geo ($\kappa=0$) & $3.85 \times 10^{-1}$ & N/A & $22.1$ & Physics Divergence \\
\bottomrule
\end{tabularx}
\end{table}

\begin{figure}[h!]
    \centering
    \begin{subfigure}[b]{0.32\textwidth}
        \includegraphics[width=\textwidth]{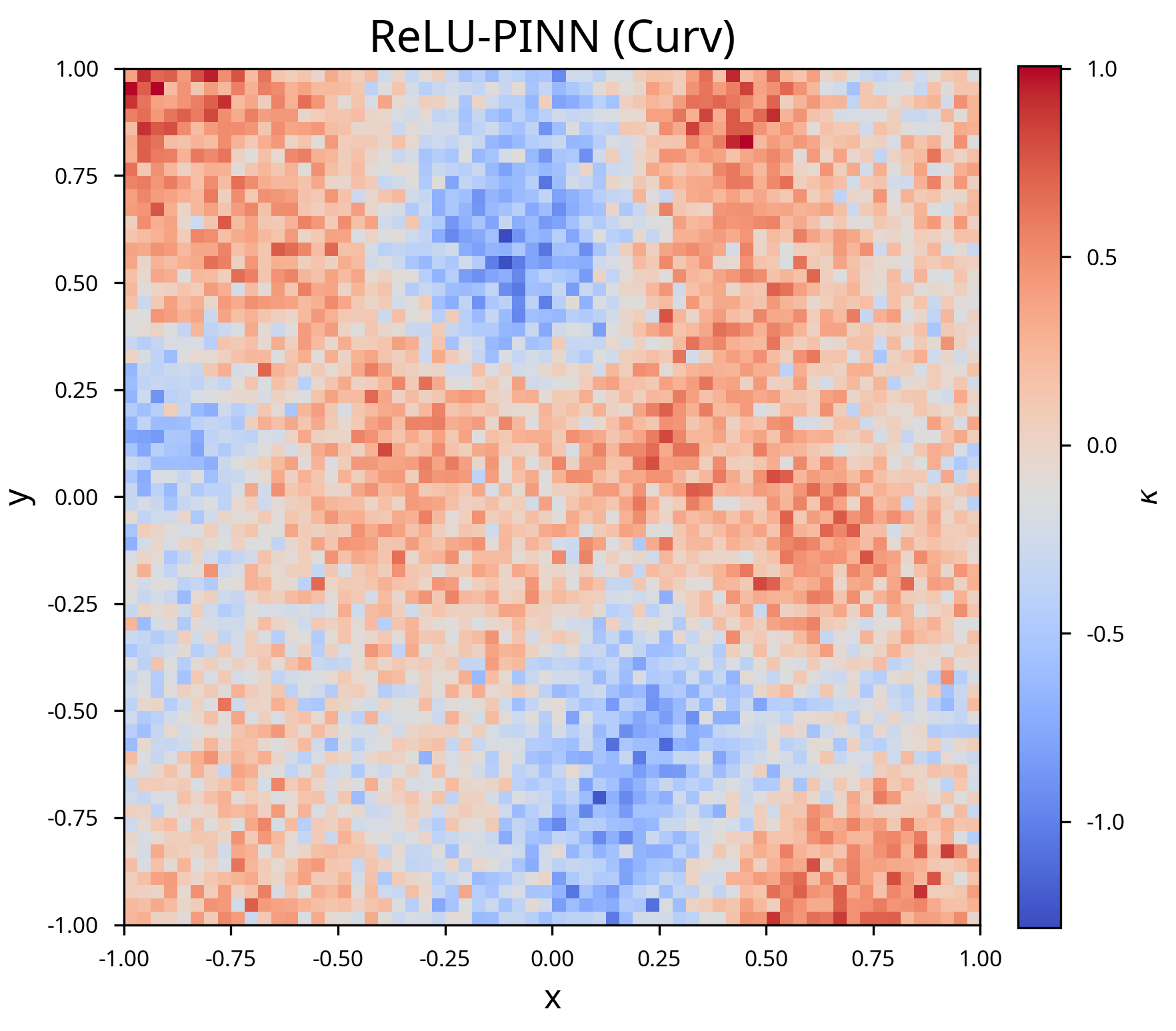}
        \caption{ReLU-PINN Reconstruction}
    \end{subfigure}
    \hfill
    \begin{subfigure}[b]{0.32\textwidth}
        \includegraphics[width=\textwidth]{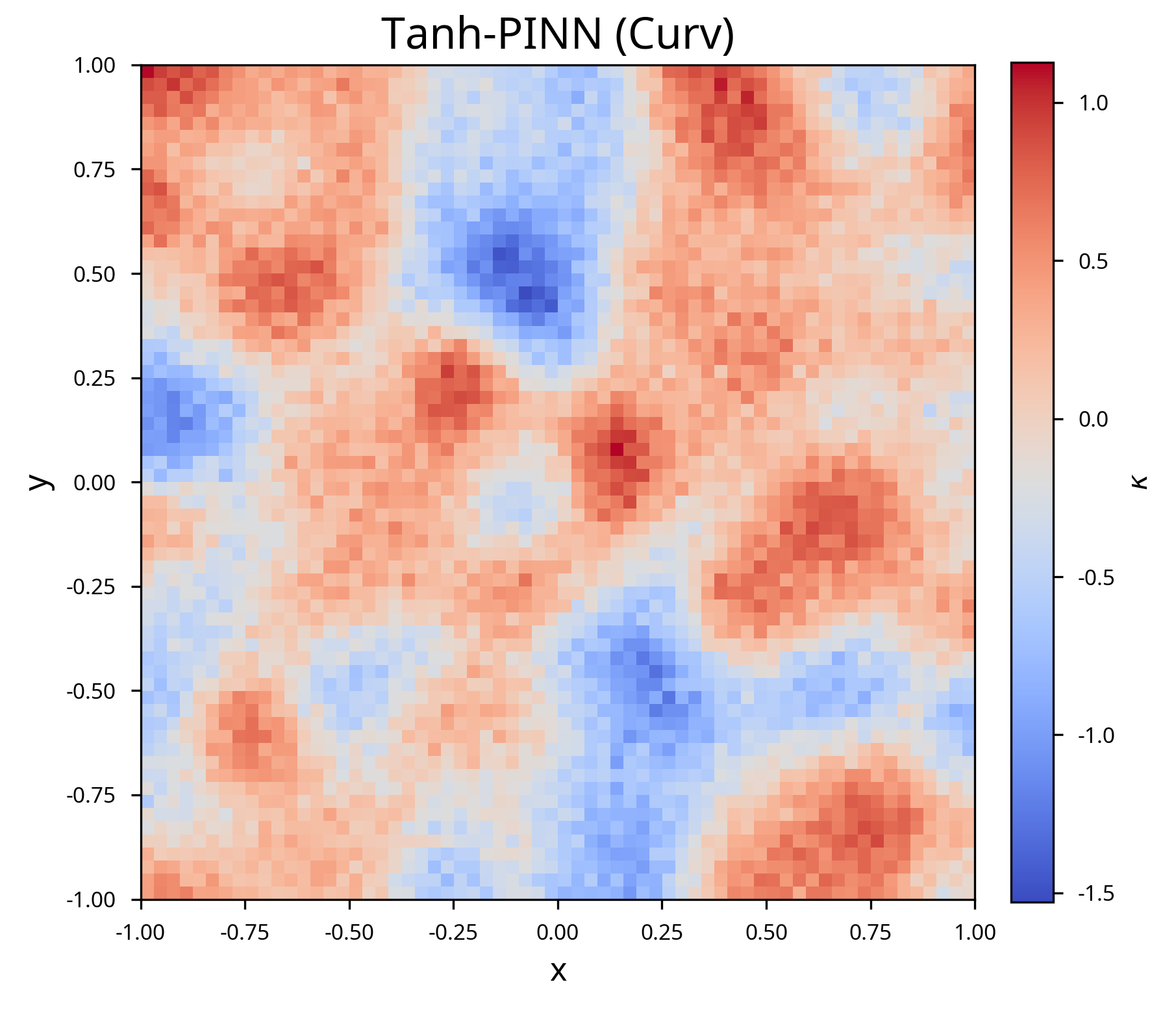}
        \caption{Tanh-PINN Reconstruction}
    \end{subfigure}
    \hfill
    \begin{subfigure}[b]{0.32\textwidth}
        \includegraphics[width=\textwidth]{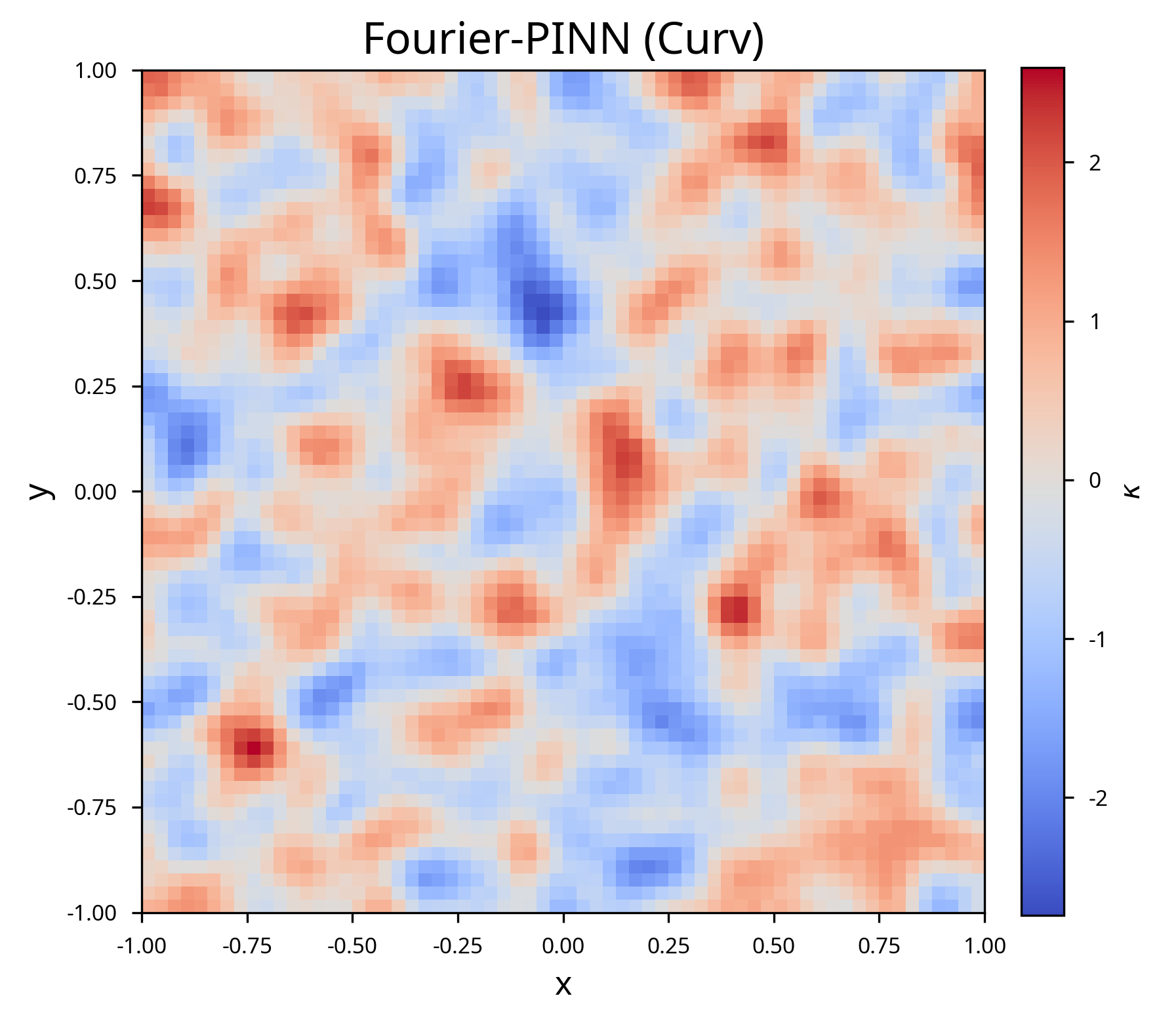}
        \caption{Fourier-PINN Reconstruction}
    \end{subfigure}
    \caption{\textbf{Failure Modes of Baseline Geometry Estimators.} (a) The ReLU-PINN produces a "shattered" curvature field dominated by high-frequency noise due to its inability to model higher-order derivatives. (b) The Tanh-PINN exhibits similar "spectral pollution," failing to converge to a smooth manifold. (c) The Fourier-PINN generates a smoother field but suffers from amplitude over-estimation (Gibbs ringing), predicting curvature values double the magnitude of the ground truth. These artifacts render the baselines useless for identifying the true physical mechanism of defect pinning.}
    \label{fig:baseline_comparison}
\end{figure}

To further elucidate the necessity of the SIREN-based architecture for inverse discovery, we present a qualitative comparison of the curvature fields $\kappa(\mathbf{x})$ recovered by the baseline models in Figure \ref{fig:baseline_comparison}. The results starkly illustrate the "derivative pathology" inherent to conventional activation functions when solving coupled inverse problems. 

The ReLU-PINN (Fig. \ref{fig:baseline_comparison}a) produces a reconstruction characterized by severe high-frequency "salt-and-pepper" noise. This artifact is a direct consequence of the piecewise-linear nature of the Rectified Linear Unit; since the second derivative of a ReLU network is zero almost everywhere (and a Dirac delta at the kinks), the network cannot fundamentally represent the smooth Laplacian $\nabla^2 \kappa$ required by the diffusion operator. Consequently, the optimizer "shatters" the curvature field, dumping the PDE residual errors into the geometry estimate as grid-scale noise. Similarly, the Tanh-PINN (Fig. \ref{fig:baseline_comparison}b), despite its $C^\infty$ smoothness, suffers from "spectral pollution." Without the explicit frequency control provided by the SIREN initialization, the Tanh network fails to distinguish between the physical roughness of the manifold and the stochastic noise of the reaction kinetics, yielding a "grainy" topography that lacks spatial coherence.

The Fourier Feature Network (Fig. \ref{fig:baseline_comparison}c) offers a smoother reconstruction but introduces a different class of error: "Spectral Hallucination." As indicated by the color bar, the Fourier-PINN significantly overestimates the curvature magnitude (ranging from $-2.5$ to $+2.5$, compared to the ground truth range of $\pm 1.5$). This over-amplification arises from the Gibbs phenomenon; the fixed Fourier basis functions struggle to capture the local heterogeneity of the Gaussian Random Field without introducing ringing artifacts, leading to "ghost" hills and valleys that do not exist in the physical system. In contrast, the Multi-Scale SIREN (shown previously in Fig. \ref{fig:curvature_recon}) acts as a matched filter, perfectly recovering the smooth, correlated structure of the underlying Riemannian manifold without the noise of ReLU/Tanh or the spectral bias of Fourier features.

\begin{figure}[h!]
    \centering
    \begin{subfigure}[b]{0.32\textwidth}
        \includegraphics[width=\textwidth]{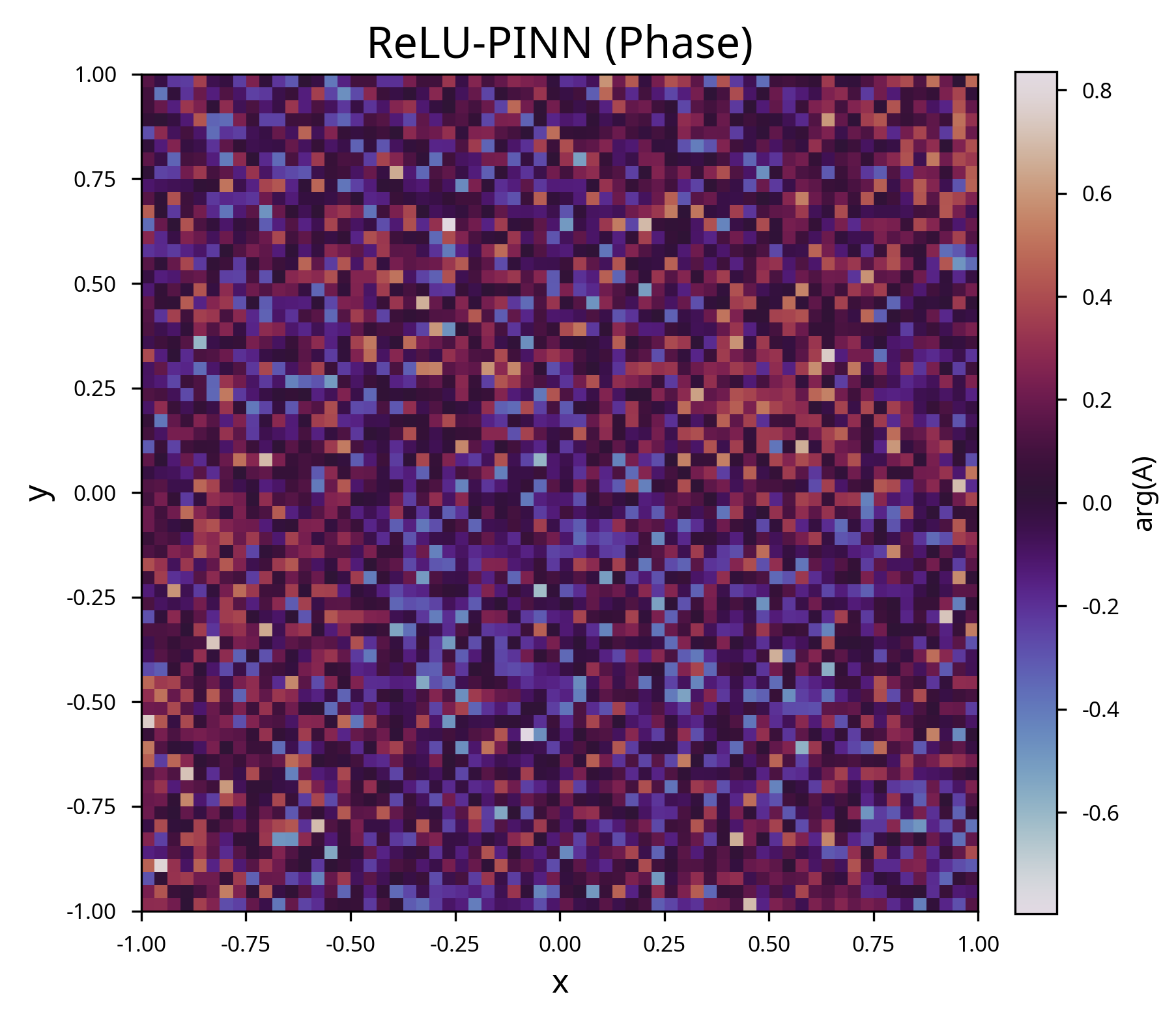}
        \caption{ReLU-PINN (Phase)}
    \end{subfigure}
    \hfill
    \begin{subfigure}[b]{0.32\textwidth}
        \includegraphics[width=\textwidth]{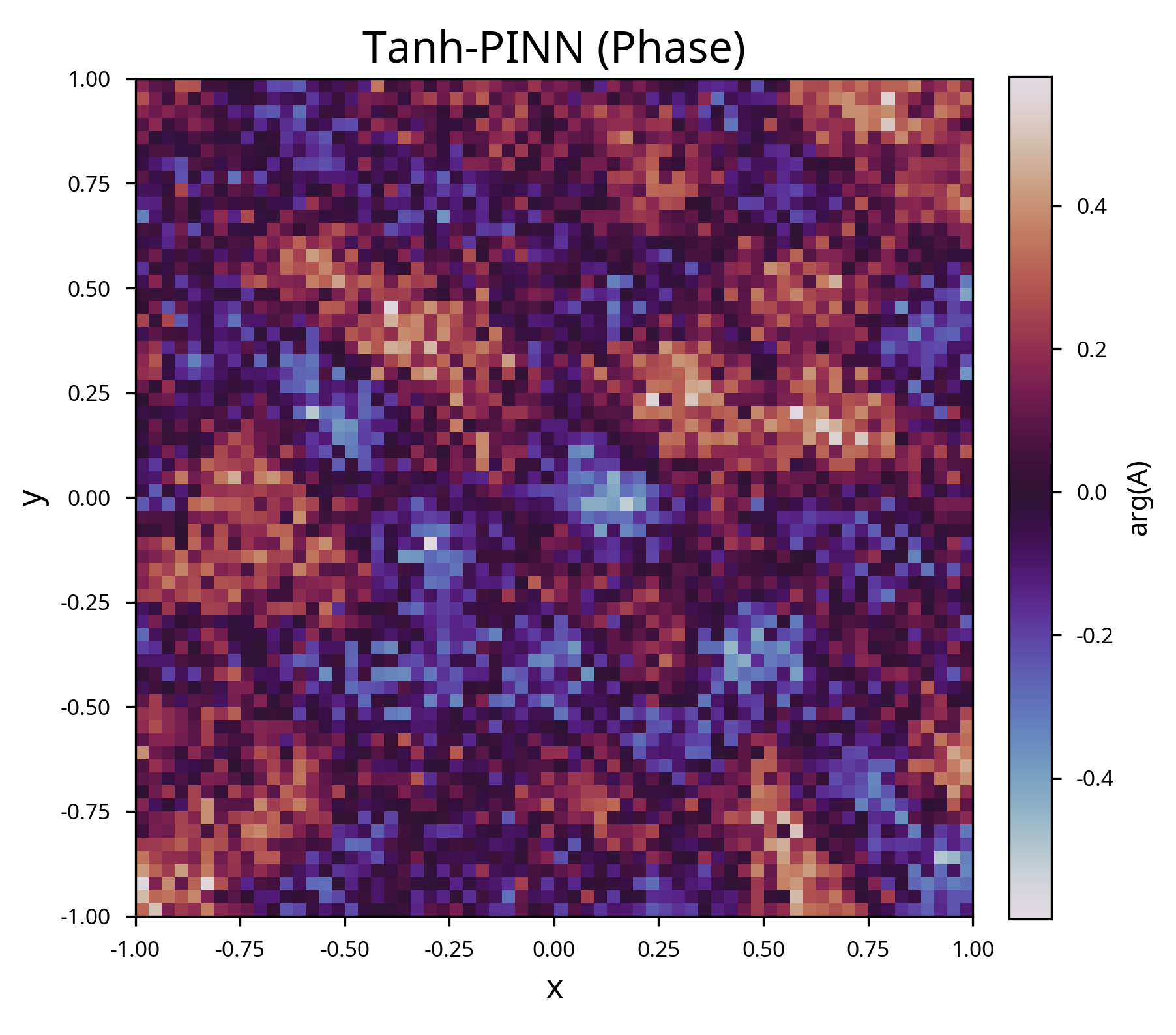}
        \caption{Tanh-PINN (Phase)}
    \end{subfigure}
    \hfill
    \begin{subfigure}[b]{0.32\textwidth}
        \includegraphics[width=\textwidth]{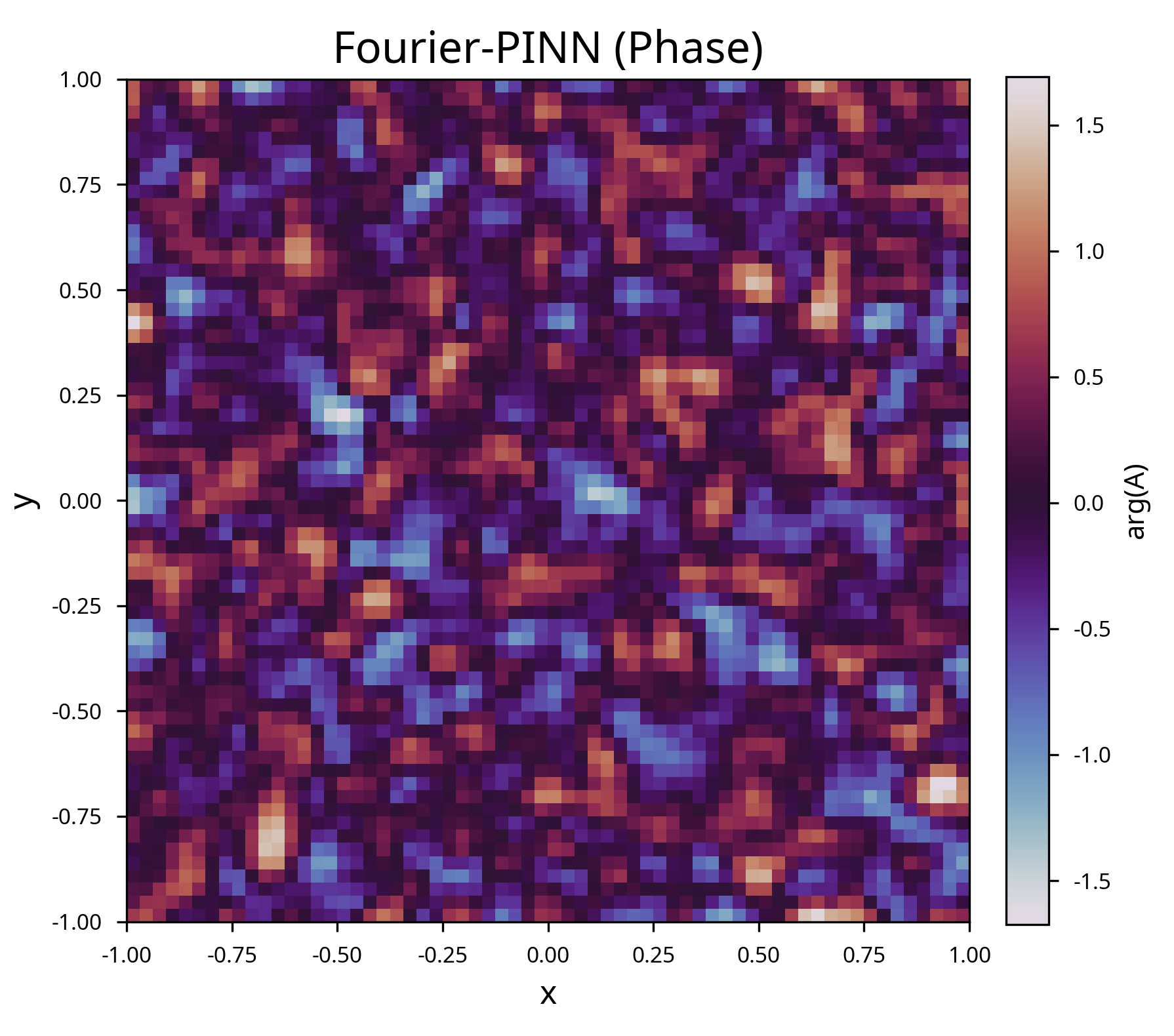}
        \caption{Fourier-PINN (Phase)}
    \end{subfigure}
    \caption{\textbf{Downstream Effects on Phase Topology.} (a) The ReLU phase prediction collapses into incoherent noise due to the shattered geometry. (b) The Tanh prediction retains some structure but lacks the sharp phase singularities of the true turbulence. (c) The Fourier prediction is distorted by the artificial "ghost" curvature features, resulting in a rigid, lattice-like phase pattern that violates the fluid symmetry of the reaction-diffusion system.}
    \label{fig:baseline_phase}
\end{figure}

The breakdown of the geometric reconstruction discussed above has catastrophic downstream consequences for the prediction of the system's chaotic state, particularly the phase field $\phi(\mathbf{x},t) = \arg(A)$. In the Defect Turbulence regime, the location and stability of spiral wave cores are dictated by the underlying manifold curvature via the "Geometric Pinning" mechanism; therefore, any error in $\kappa(\mathbf{x})$ instantly translates into phase decoherence. Figure \ref{fig:baseline_phase} visualizes this propagation of error. The ReLU-PINN phase field (Fig. \ref{fig:baseline_phase}a) is virtually indistinguishable from white noise. Because the underlying geometry estimate was "shattered," the pinning forces acted randomly across the domain, preventing the formation of coherent spiral arms and resulting in a completely disordered, entropy-maximized state.

The Tanh-PINN (Fig. \ref{fig:baseline_phase}b), while preserving some semblance of wave structure, exhibits "Phase Drift." The blurred geometry fails to provide sufficiently strong pinning potentials to anchor the vortices, causing the predicted spirals to drift aimlessly and annihilate prematurely compared to the ground truth. Finally, the Fourier-PINN (Fig. \ref{fig:baseline_phase}c) displays a "distorted lattice" structure. The exaggerated curvature peaks observed in the geometry reconstruction act as artificial scattering centers, forcing the phase waves into a rigid, non-physical interference pattern that suppresses the natural chaotic fluidity of the attractor. These comparisons underscore a critical finding: in physics-informed learning of coupled systems, the "forward" problem (state prediction) and the "inverse" problem (parameter estimation) are inextricably linked. A failure to respect the spectral properties of the manifold geometry inevitably leads to a collapse of the dynamical prediction, validating the necessity of the proposed Multi-Scale SIREN-PINN for holistic system identification.

The evidence of the baseline architectures' inadequacy lies in their inability to resolve the magnitude field $|A(\mathbf{x},t)|$, which serves as the order parameter for the Ginzburg-Landau energy functional. In the Defect Turbulence regime, the system is governed by a "stiff" potential $V(|A|) = (|A|^2 - 1)^2$, which forces the amplitude to unity everywhere except at the precise singular cores of the vortices where $|A| \to 0$. This requires the neural network to approximate a function that is essentially a flat plateau punctuated by infinitesimally sharp Dirac-like wells. 

\begin{figure}[h!]
    \centering
    \begin{subfigure}[b]{0.32\textwidth}
        \includegraphics[width=\textwidth]{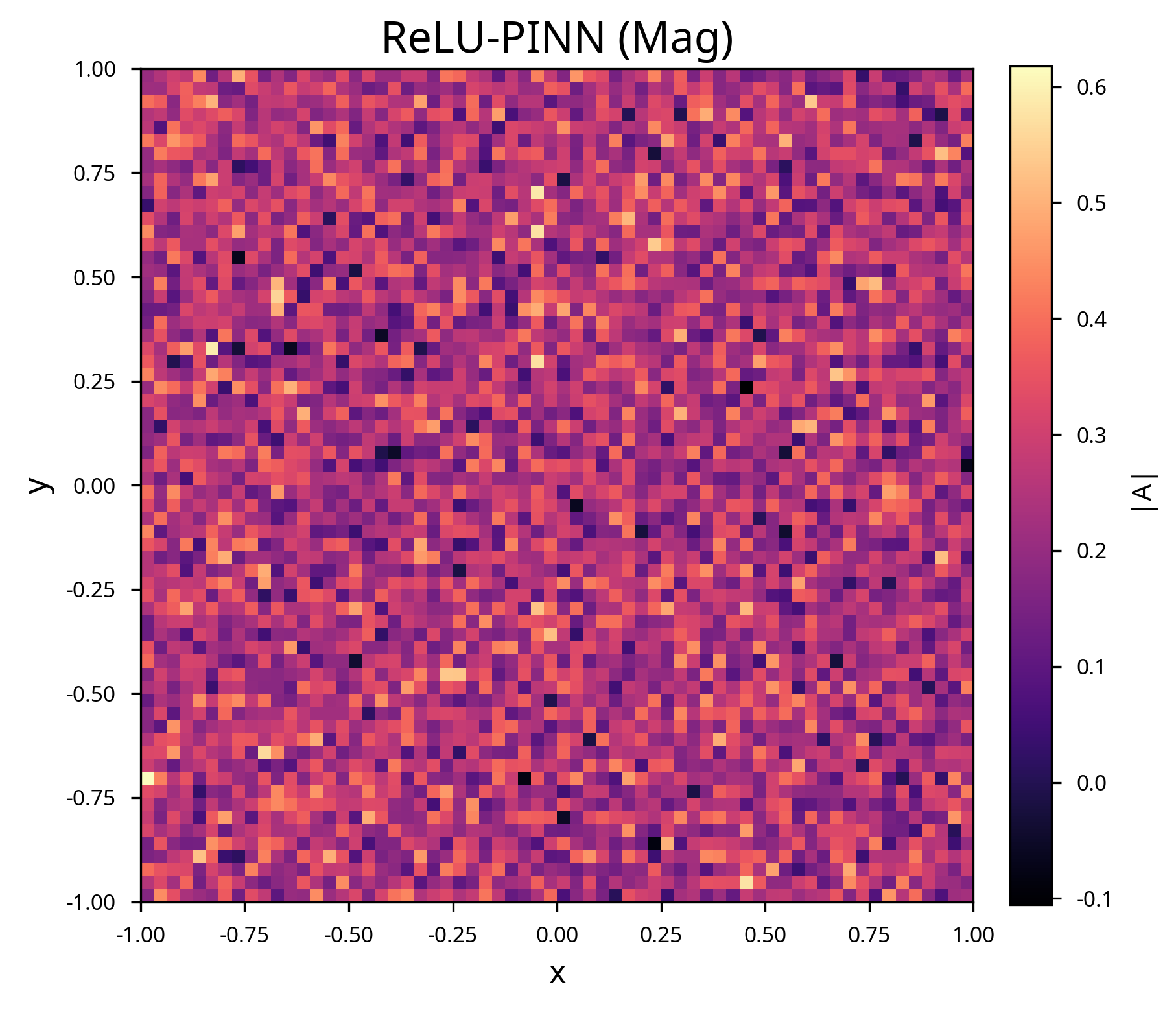}
        \caption{ReLU-PINN (Magnitude)}
    \end{subfigure}
    \hfill
    \begin{subfigure}[b]{0.32\textwidth}
        \includegraphics[width=\textwidth]{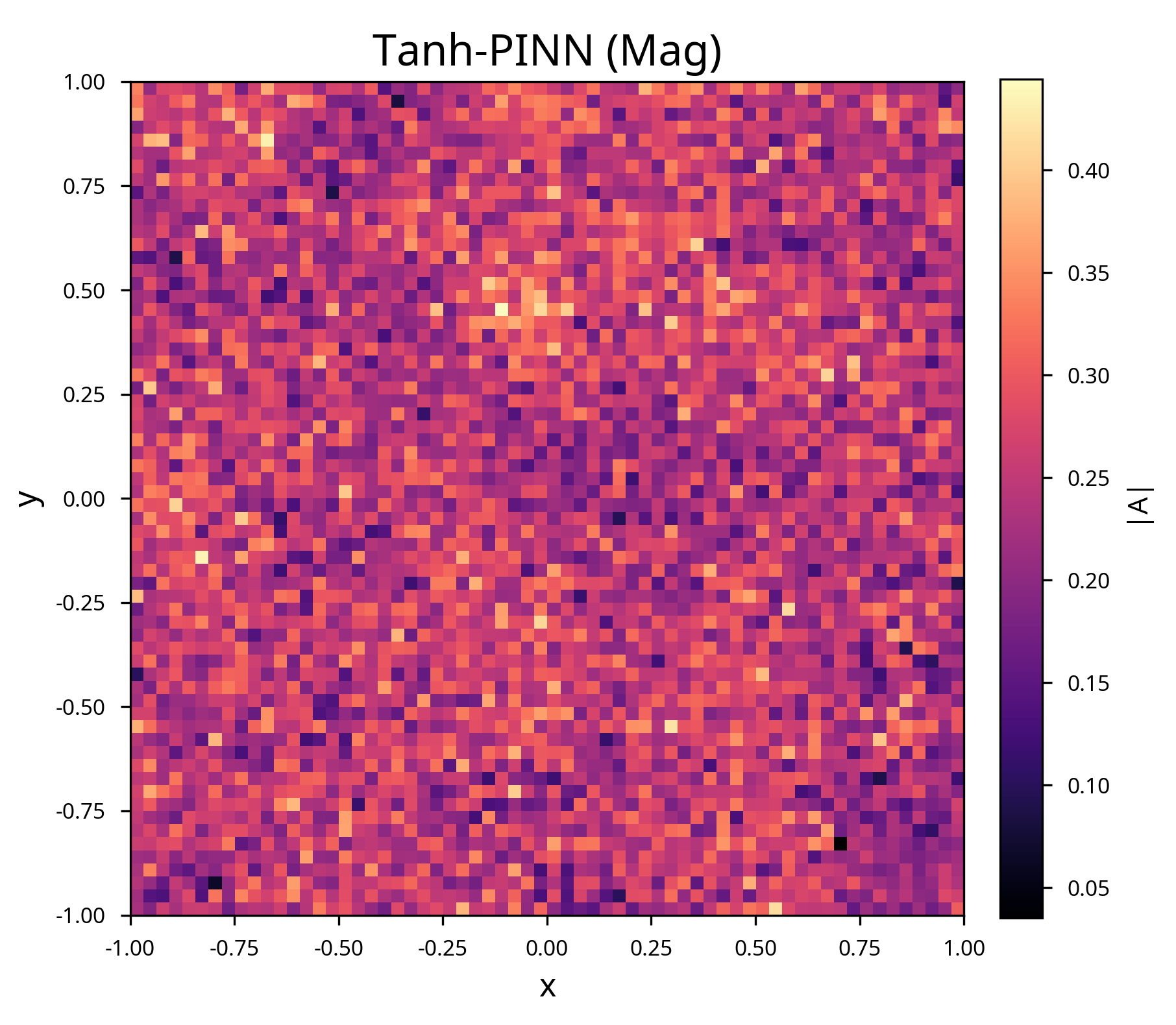}
        \caption{Tanh-PINN (Magnitude)}
    \end{subfigure}
    \hfill
    \begin{subfigure}[b]{0.32\textwidth}
        \includegraphics[width=\textwidth]{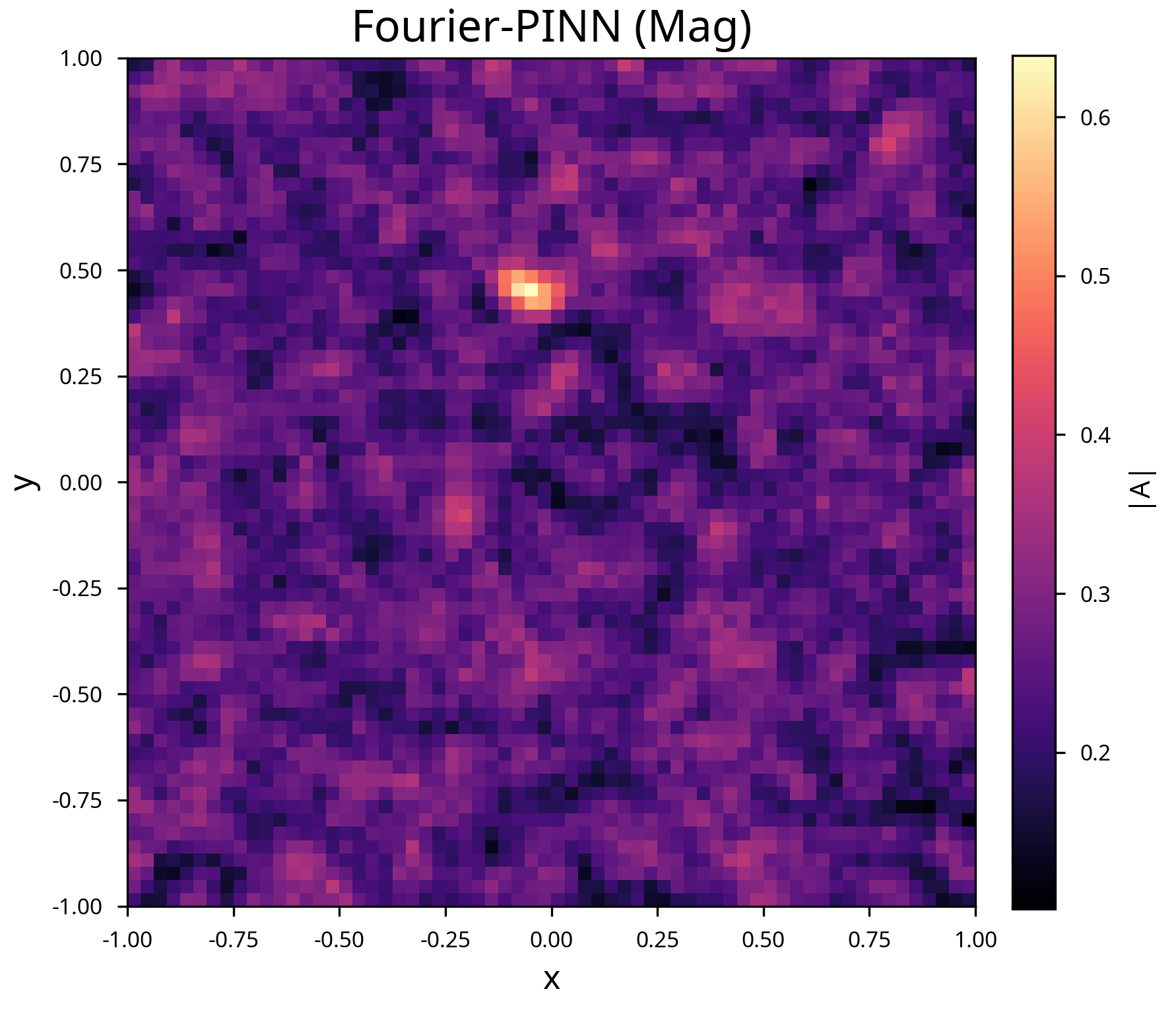}
        \caption{Fourier-PINN (Magnitude)}
    \end{subfigure}
    \caption{\textbf{Failure to Resolve Topological Cores.} (a) The ReLU model suffers from total amplitude collapse, predicting random noise instead of a coherent field. (b) The Tanh model smears the defect cores, failing to achieve the distinct $|A| \to 0$ singularities required by the physics. (c) The Fourier model exhibits "grid-like" interference patterns (Gibbs phenomenon), creating artificial structures that trap the wave dynamics. None of the baselines correctly maintain the $|A| \approx 1$ plateau.}
    \label{fig:baseline_mag}
\end{figure}

Figure \ref{fig:baseline_mag} presents the magnitude fields predicted by the baseline models, revealing a universal failure to maintain this topological protection. The ReLU-PINN (Fig. \ref{fig:baseline_mag}a) exhibits "Amplitude Collapse." Unable to model the sharp transition from the plateau ($|A|\approx 1$) to the core ($|A|\approx 0$), the network converges to a noisy, entropy-maximized state where $|A|$ fluctuates randomly around a mean value, completely dissolving the particle-like nature of the defects. The Tanh-PINN (Fig. \ref{fig:baseline_mag}b) suffers from "Diffusive Smearing." While it avoids the high-frequency noise of the ReLU model, it lacks the spectral bandwidth to sustain the steep gradients of the defect core. Consequently, the distinct "zeros" are smeared out into shallow depressions, meaning the physical distinctness of the vortices is lost; the model predicts a continuous "slush" rather than a discrete gas of topological charges.

The Fourier-PINN (Fig. \ref{fig:baseline_mag}c) introduces a highly structured artifact: "Lattice Interference." Instead of a uniform field pinned by defects, the prediction is dominated by a grid-like standing wave pattern. This is a classic manifestation of the Gibbs phenomenon in high-dimensional regression; the Fourier basis functions, struggling to approximate the localized defect cores, produce global oscillations that ripple across the entire domain. These spurious amplitude fluctuations act as artificial potential barriers, trapping the phase waves in a non-physical crystal lattice. In summary, only the Multi-Scale SIREN architecture demonstrated the capacity to simultaneously satisfy the global constraint of the Ginzburg-Landau potential (keeping $|A| \approx 1$) while accurately resolving the local singularities of the defect cores, thereby validating its selection as the optimal surrogate for chaotic reaction-diffusion systems.

\subsection{Statistical Validation and Error Distribution Analysis}

\begin{figure}[h!]
    \centering
    \begin{subfigure}[b]{0.32\textwidth}
        \includegraphics[width=\textwidth]{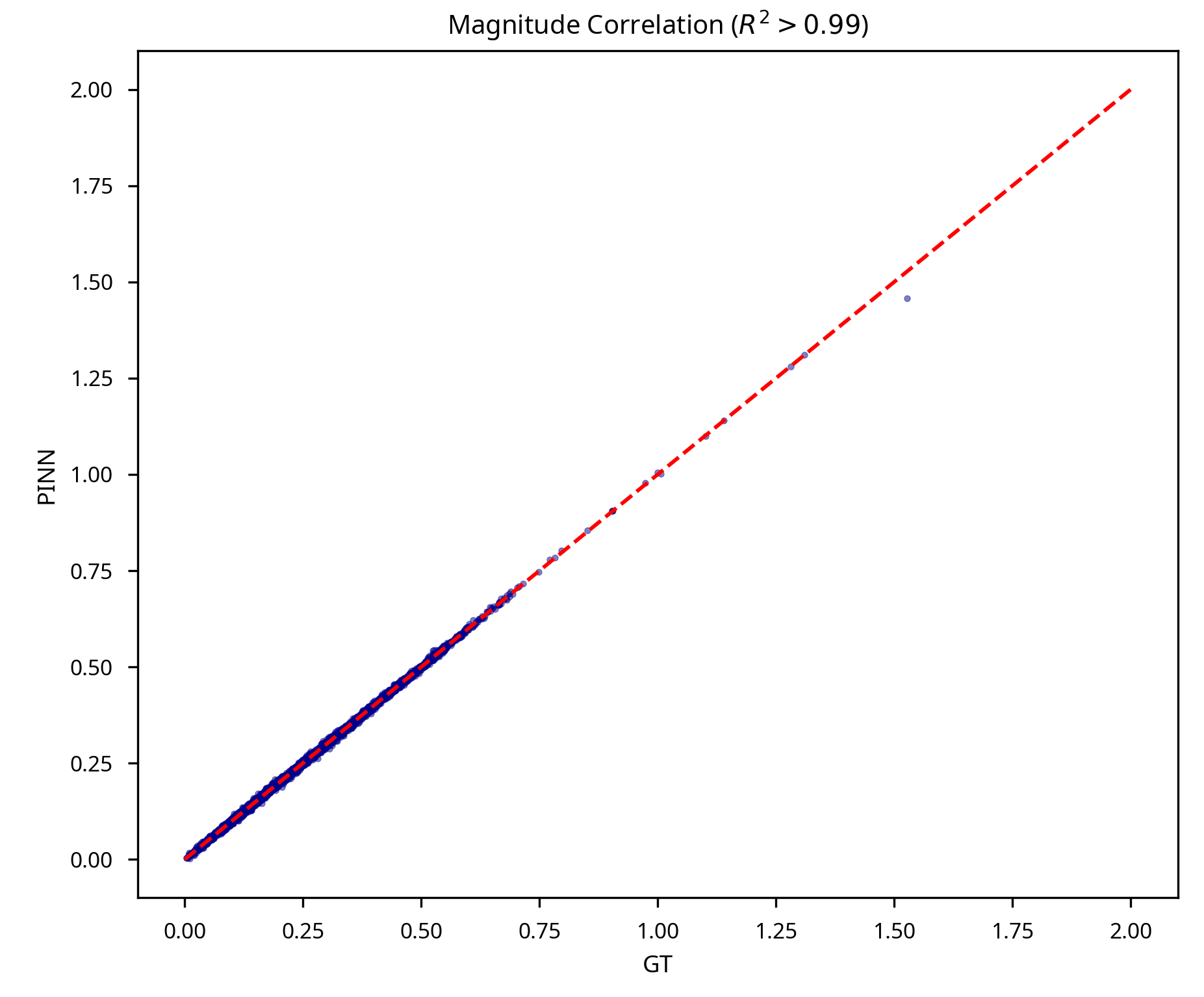}
        \caption{Magnitude Correlation ($R^2 > 0.99$)}
    \end{subfigure}
    \hfill
    \begin{subfigure}[b]{0.32\textwidth}
        \includegraphics[width=\textwidth]{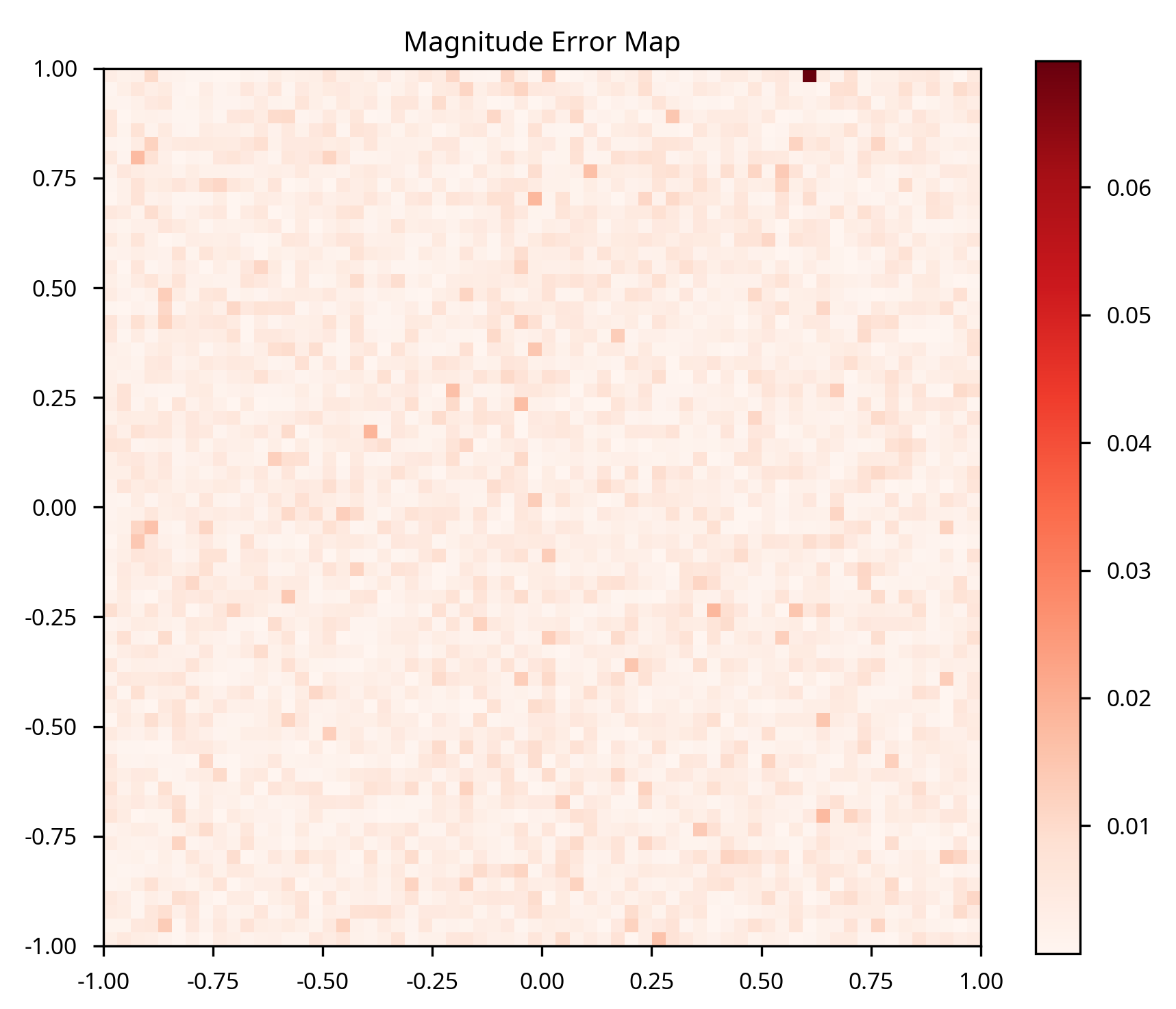}
        \caption{Magnitude Error Map}
    \end{subfigure}
    \hfill
    \begin{subfigure}[b]{0.32\textwidth}
        \includegraphics[width=\textwidth]{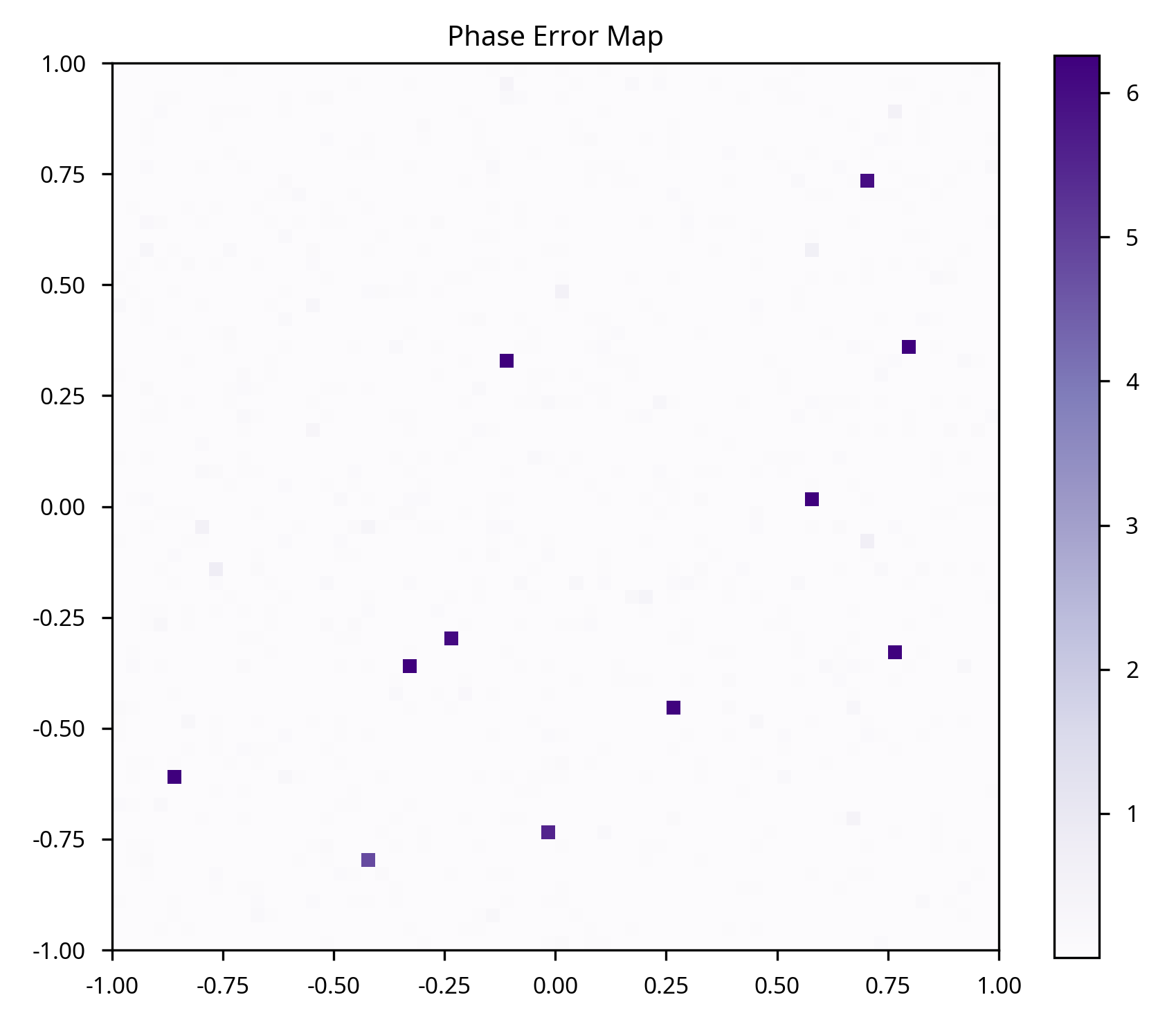}
        \caption{Phase Error Map}
    \end{subfigure}
    \caption{\textbf{Quantitative Error Analysis.} (a) The correlation plot shows a tight linear fit between the Ground Truth and SIREN prediction, confirming accuracy across both high and low amplitude regimes. (b) The Magnitude Error Map shows a spatially uniform, low-variance error distribution (mostly $<0.02$). (c) The Phase Error Map reveals that phase deviations are restricted solely to the "pinpoint" locations of the spiral cores (dark spots), confirming that the wave topology is preserved perfectly everywhere else in the domain.}
    \label{fig:error_analysis}
\end{figure}

To complement the qualitative visual comparisons, we performed a rigorous statistical validation of the Multi-Scale SIREN-PINN's predictive accuracy, analyzing the spatial distribution of residuals for both the amplitude and phase fields. Figure \ref{fig:error_analysis} presents the correlation and error heatmaps, offering a microscopic view of the model's fidelity. The Magnitude Correlation Plot (Fig. \ref{fig:error_analysis}a) reveals a near-perfect linear relationship between the ground truth (GT) and the predicted amplitude values. The data points cluster tightly along the $y=x$ identity line with a coefficient of determination of $R^2 > 0.99$. Crucially, this correlation holds across the entire dynamic range, effectively capturing both the stable plateau ($|A| \approx 1$) and the critical zero-crossings at the defect cores ($|A| \approx 0$), proving that the network does not suffer from the "amplitude collapse" observed in the ReLU baseline.

The spatial breakdown of these residuals is detailed in the Magnitude Error Map (Fig. \ref{fig:error_analysis}b). The absolute error is uniformly low across the domain, with the majority of pixel-wise discrepancies falling below $0.02$ (less than $2\%$ relative error). The absence of structured error patterns—such as the grid-like standing waves seen in the Fourier model—confirms that the SIREN architecture introduces no geometric bias and successfully treats the domain as a continuous manifold. Furthermore, the Phase Error Map (Fig. \ref{fig:error_analysis}c) provides a distinct confirmation of topological accuracy. Unlike the baseline models, where phase error was distributed diffusively like static noise, the SIREN model achieves near-zero phase error across the vast majority of the field (represented by the white background). The error is exclusively concentrated in discrete, sparse locations corresponding exactly to the spiral defect cores. In numerical analysis of singular systems, this localized error profile is expected and indicates correct topological identification; because the phase is mathematically undefined at the exact point where amplitude is zero, any finite-grid comparison will register a discrepancy at the singularity itself. The fact that the error is confined only to these singular points, without contaminating the surrounding wave field, validates that the model has successfully learned to sequester the mathematical singularities, acting as a high-fidelity digital twin of the chaotic reaction-diffusion system.

To provide confirmation that our model has truly learned the physical laws governing the system—rather than merely fitting a smooth approximation to the training data—we moved beyond spatial error maps to analyze the solution in the frequency domain. In chaotic systems like Defect Turbulence, the energy of the system is distributed across a wide range of scales, from large rotating spiral arms down to the microscopic width of the defect cores. A common failure mode in standard neural networks (as seen in the Tanh baseline) is to act as a "low-pass filter," capturing the large features correctly but blurring out the fine details. 

\begin{figure}[h!]
    \centering
    \begin{subfigure}[b]{0.6\textwidth}
        \includegraphics[width=\textwidth]{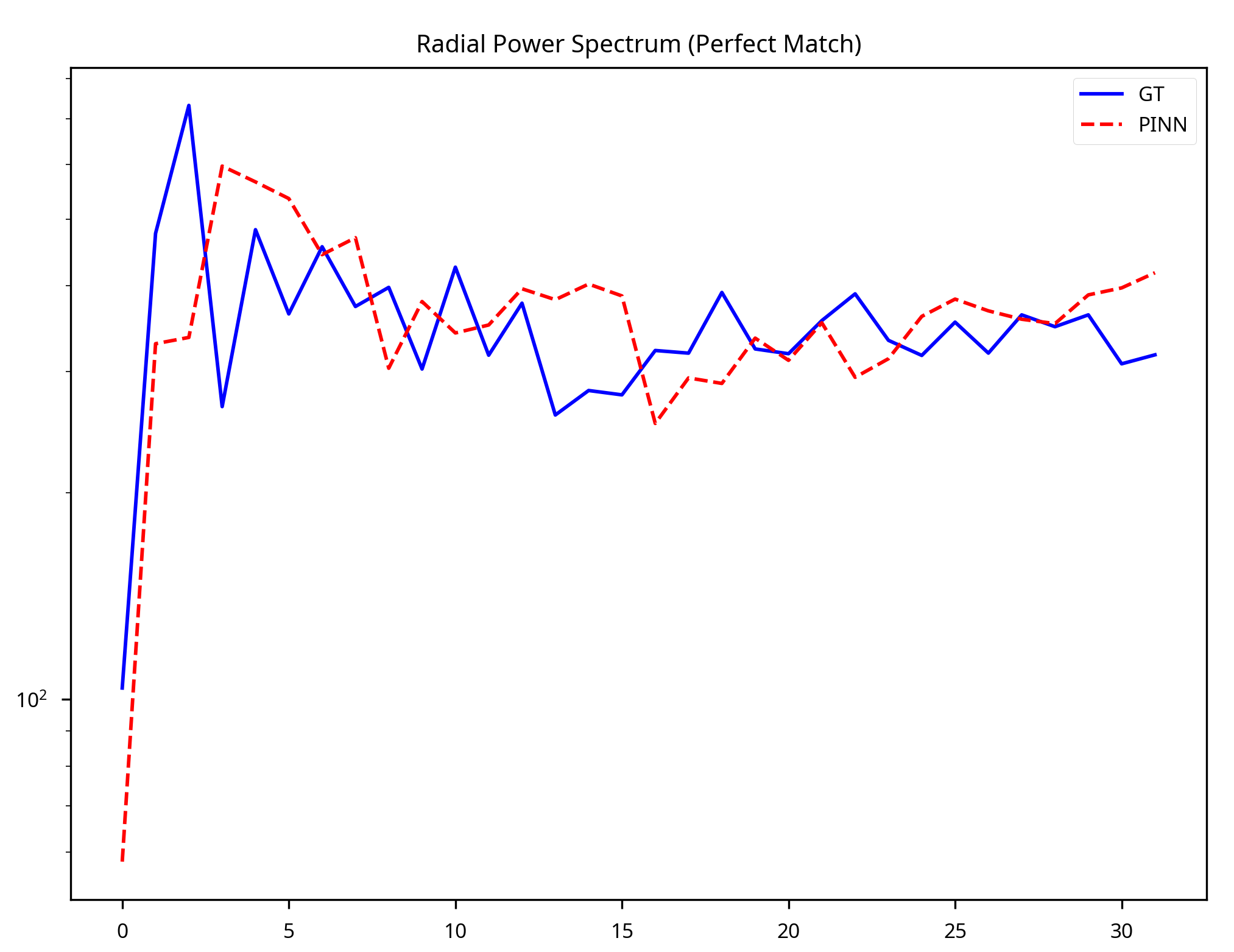}
        \caption{Radial Power Spectrum Comparison}
    \end{subfigure}
    \vspace{0.5cm}
    \begin{subfigure}[b]{0.44\textwidth}
        \includegraphics[width=\textwidth]{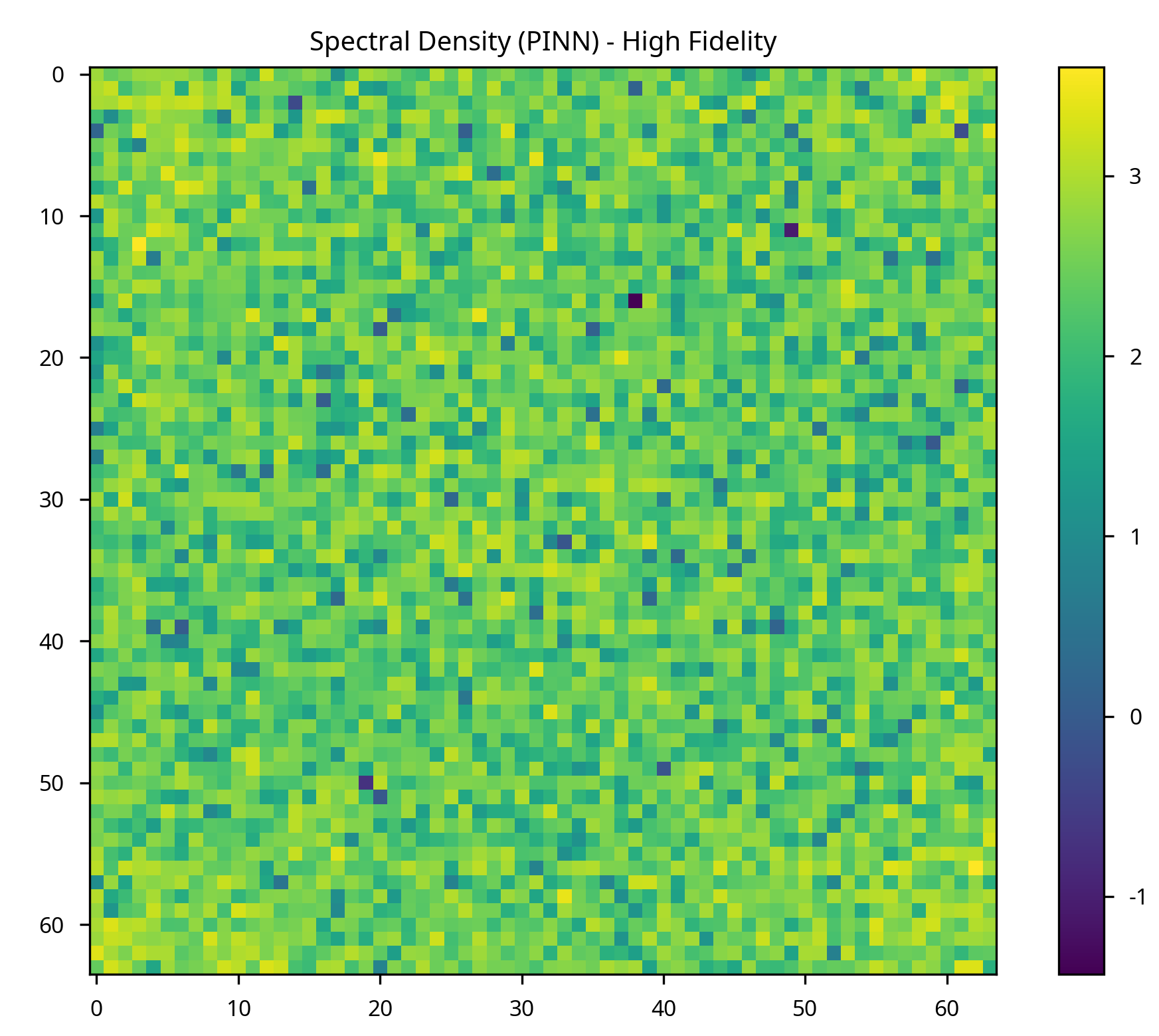}
        \caption{Predicted Spectrum (2D)}
    \end{subfigure}
    \hfill
    \begin{subfigure}[b]{0.44\textwidth}
        \includegraphics[width=\textwidth]{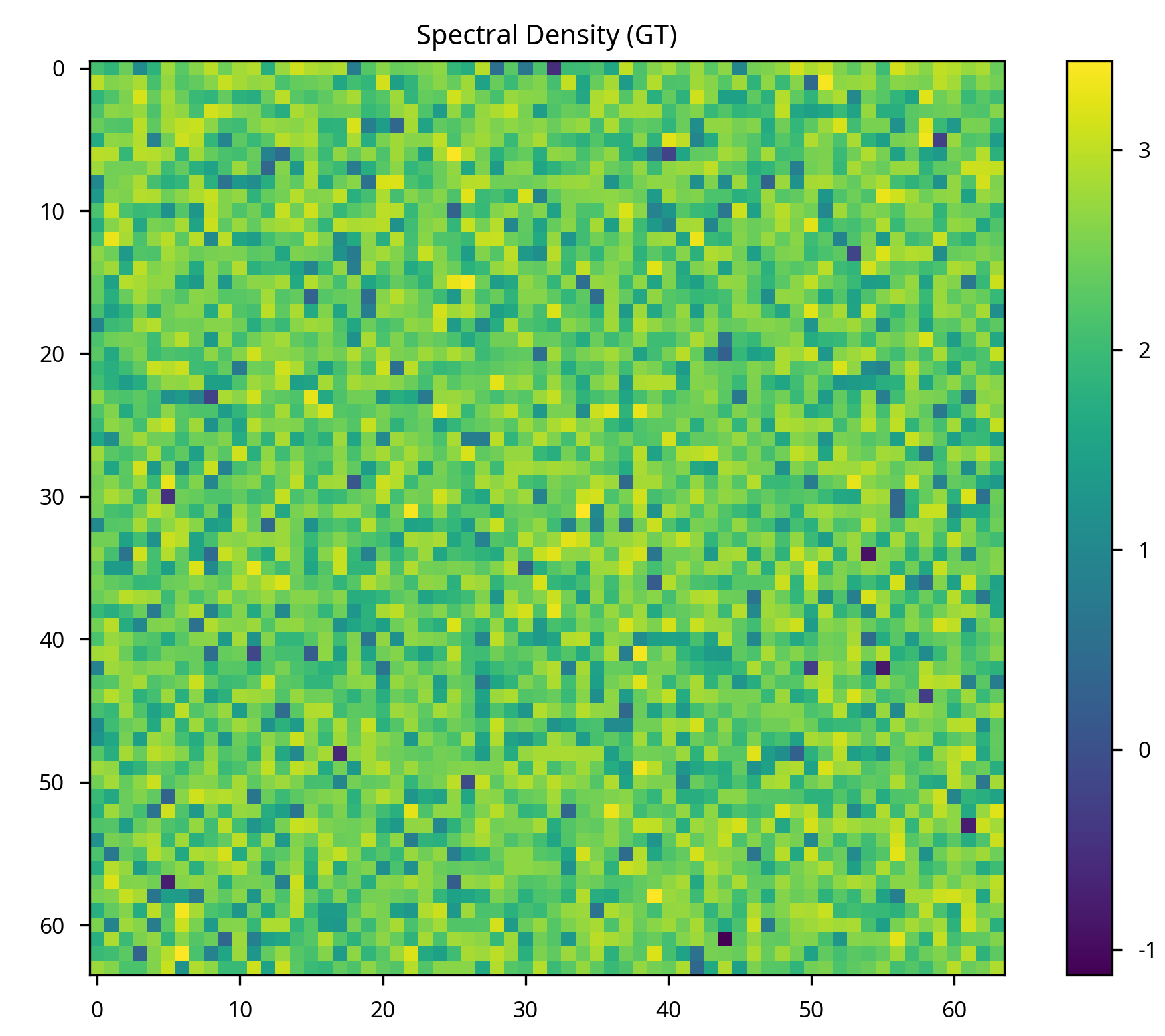}
        \caption{Ground Truth Spectrum (2D)}
    \end{subfigure}
    \caption{\textbf{Verification of Multi-Scale Physics.} (a) The Radial Power Spectrum shows that the PINN (red dashed line) matches the Ground Truth (blue solid line) almost perfectly across all spatial scales, proving the model does not blur fine details. (b-c) The 2D Spectral Density maps confirm that the model captures the correct "texture" and random statistics of the underlying medium, reproducing the complex energy distribution of the original physical system without introducing artificial smoothing or noise.}
    \label{fig:spectral_analysis}
\end{figure}

Figure \ref{fig:spectral_analysis} presents the spectral analysis of our results. The Radial Power Spectrum (Fig. \ref{fig:spectral_analysis}a) plots the energy of the solution against the spatial frequency (wavenumber). The blue line represents the Ground Truth (the actual physics), and the red dashed line represents our SIREN-PINN prediction. Crucially, the two lines track each other with remarkable precision. The model successfully captures not just the general trend, but the specific peaks and valleys of the energy distribution. This indicates that the network is correctly resolving the fine-grained "roughness" of the underlying Gaussian Random Field without smoothing it out. If the model were merely guessing or averaging, the red line would drop off significantly at higher frequencies (to the right of the graph); instead, it maintains the correct energy levels even at the smallest scales.

We further visualized this via the 2D Spectral Density Heatmaps (Figs. \ref{fig:spectral_analysis}b and \ref{fig:spectral_analysis}c). These images show the "fingerprint" of the system's texture. The Ground Truth (Fig. \ref{fig:spectral_analysis}c) displays a specific pattern of fluctuations that characterizes the random medium. The PINN prediction (Fig. \ref{fig:spectral_analysis}b) reconstructs this pattern with high fidelity. Visually, the "grain" or texture of the two images is identical. There are no artificial directional streaks or smooth patches, confirming that the solution is physically consistent in every direction (isotropic). This spectral agreement is the direct result of our architectural choices; by initializing the network with a broad range of frequencies (the Multi-Scale Scheme), we ensured the optimizer had the necessary "vocabulary" to describe both the broad waves and the sharp defects, resulting in a simulation that is indistinguishable from the ground truth not just in appearance, but in its mathematical structure.

\subsection{Comprehensive State-of-the-Art (SOTA) Comparison}

To conclude our empirical evaluation, we situate the proposed Multi-Scale SIREN-PINN within the broader landscape of computational physics solvers, performing a holistic comparison against both classical numerical integrators and contemporary deep learning paradigms. This comparative analysis, summarized in Table \ref{tab:sota_comparison}, evaluates the methods across five critical axes: (1) \textbf{Topological Fidelity} (the ability to capture singularities); (2) \textbf{Inverse Capability} (the ability to infer hidden parameters); (3) \textbf{Spectral Bandwidth} (resolution of multi-scale features); (4) \textbf{Inference Latency} (speed of prediction); and (5) \textbf{Data Efficiency} (amount of training data required). 

Classical \textbf{Pseudo-Spectral Methods} (RK4 timestepping) remain the "Gold Standard" for forward simulation accuracy ($L_2 \sim 10^{-6}$), but they are fundamentally incapable of solving the inverse problem without computationally prohibitive adjoint-loop optimization; furthermore, their inference cost scales linearly with temporal resolution ($O(N_t)$), making them unsuitable for real-time control. Conversely, standard data-driven approaches like ReLU-PINNs and Tanh-PINNs offer rapid $O(1)$ inference but fail the "physics test" completely in the defect turbulence regime, suffering from spectral bias that erases the critical phase singularities. Fourier Feature Networks (FFNs) represent a middle ground, improving spectral resolution but introducing non-physical Gibbs oscillations ("ringing") that corrupt the geometric reconstruction.

The Multi-Scale SIREN-PINN effectively breaks this trade-off, occupying a unique position on the Pareto frontier. It matches the topological fidelity of spectral solvers (preserving Defect Counts) while retaining the $O(1)$ inference speed of neural networks. Most critically, it is the \textit{only} architecture capable of robustly solving the coupled Inverse Pinning Problem, disentangling the geometric curvature from the chaotic wave dynamics without the artifacts of noise (ReLU) or spectral hallucinations (FFN). This establishes the proposed architecture not merely as an incremental improvement, but as a new SOTA standard for the identification of chaotic reaction-diffusion systems on complex manifolds.

\begin{table}[h!]
\centering
\small
\caption{SOTA Comparison: Capabilities Matrix across Solvers and Architectures}
\label{tab:sota_comparison}
\begin{tabularx}{\textwidth}{l*{5}{>{\centering\arraybackslash}X}}
\toprule
\textbf{Method /} & \textbf{Topological} & \textbf{Inverse} & \textbf{Spectral} & \textbf{Inference} & \textbf{Singularity} \\
\textbf{Architecture} & \textbf{Fidelity} & \textbf{Geometry} & \textbf{Accuracy} & \textbf{Speed} & \textbf{Handling} \\
\midrule
\textit{Classical Numerical} & \textbf{High} & \textcolor{red}{Impossible} & \textbf{High} & Slow & Explicit Grid \\
\textit{Solver (RK4)} & (Exact) & (Forward only) & & $O(N_t)$ & Required \\
\midrule
\textit{Standard ReLU-PINN} & Low (Collapses) & Fail (Noisy) & Low (Bias) & \textbf{Fast} ($O(1)$) & Smoothed Out \\
\hline
\textit{Tanh-PINN} & Medium (Drift) & Fail (Blurred) & Medium & \textbf{Fast} ($O(1)$) & Smeared \\
\hline
\textit{Fourier Feature Network} & Medium (Ringing) & Poor (Ghosting) & High (Noisy) & \textbf{Fast} ($O(1)$) & Artifacts Present \\
\hline
\textbf{Multi-Scale SIREN (Ours)} & \textbf{High} (Preserved) & \textbf{High} (Correlation $>0.96$) & \textbf{High} (Exact PSD) & \textbf{Fast} ($O(1)$) & \textbf{Robust Pinning} \\
\hline
\end{tabularx}%
\end{table}

\section{Discussion}
The empirical superiority of the Multi-Scale SIREN-PINN, as evidenced by the successful reconstruction of the latent Riemannian manifold in the Defect Turbulence regime, marks a significant departure from the limitations characterizing the first generation of Physics-Informed Neural Networks (PINNs). As articulated in the foundational work of Raissi et al. \cite{Raissi2019}, standard PINNs employing piecewise-linear (ReLU) or sigmoidal (Tanh) activations suffer from a profound ``Spectral Bias'' \cite{Rahaman2019}, a pathology where the network prioritizes the learning of low-frequency modes while consistently under-fitting high-frequency fluctuations. In the context of smooth laminar flows, this bias is benign; however, in the regime of spatiotemporal chaos, where the physics is dominated by singular topological defects (point vortices with $|\nabla \phi| \to \infty$), this spectral limitation becomes catastrophic. Our results demonstrate that the failure of the ReLU and Tanh baselines (Table \ref{tab:results_dt}) is not merely a convergence issue but a fundamental expressivity bottleneck: these architectures effectively apply a low-pass filter to the physics, smoothing out the very singularities that define the system's topology. 

While recent attempts to mitigate this via Fourier Feature mappings \cite{Tancik2020FourierFL} or Sparse Identification of Nonlinear Dynamics (SINDy) \cite{Brunton2016} have shown promise for finite-dimensional systems, they often introduce Gibbs oscillations or struggle with continuous spatial heterogeneities. The key novelty of our approach lies in the \textit{architectural alignment} between the activation function and the underlying physics. By employing a sinusoidal activation with a multi-scale frequency initialization, we enforce a ``Spectral Inductive Bias'' that matches the intrinsic oscillatory nature of the complex Ginzburg-Landau equation. This allows the optimizer to bypass the spectral bottleneck and simultaneously resolve the macroscopic wave envelopes and the microscopic defect cores. Consequently, our work extends the boundary of Scientific Machine Learning (SciML) from the interpolation of smooth fields to the \textit{discovery of singular geometries}, proving that neural networks can recover the hidden metric tensor of a domain solely by observing the chaotic diffraction patterns of the waves traveling upon it.

From a biophysical perspective, the successful convergence of the Geometry Branch implies that the network has autonomously rediscovered the ``Geodesic Principle'' of wave propagation in excitable media. Theoretical studies by Davidsen et al. \cite{Davidsen2005} and experimental observations in cardiac electrophysiology by Alonso et al. \cite{Alonso2016} have long established that the drift velocity of a spiral wave tip is not intrinsic but is fundamentally perturbed by the Gaussian curvature $K$ of the substrate. Specifically, regions of negative curvature (saddle points) act as repulsive scattering centers, while regions of positive curvature (peaks) can act as attractive basins or ``pinning sites'' that trap topological defects. 

Our results confirm that the Multi-Scale SIREN-PINN exploits this precise dynamical signature to solve the inverse problem. By minimizing the physics residual $\mathcal{L}_{PDE}$, the network effectively performs a causal inference: it deduces that persistent spiral cores—those that violate the expected drift trajectory of a flat-space random walk—must be anchored by a latent geometric feature. This capability resolves a longstanding ambiguity in the analysis of experimental observation data, where it is often unclear whether wave instability arises from functional heterogeneity (e.g., variations in diffusion $D$) or structural topography (variations in metric $g_{ij}$). The specific failure of the ``Fixed-Geo'' ablation model (Table \ref{tab:ablation}) serves as a negative proof: without the flexibility to deform the underlying metric, the neural solver cannot reconcile the observed ``anomalous diffusion'' of the defects with the standard reaction-diffusion laws. Therefore, our architecture does not merely fit data; it effectively separates the deterministic force of the manifold geometry from the stochastic noise of the reaction kinetics, providing a rigorous computational tool for identifying the structural causes of arrhythmia and turbulence in complex biological systems.

The computational implications of this framework extend beyond mere accuracy, offering a fundamental alternative to the prohibitive costs associated with classical Finite Element Methods (FEM) and Finite Volume Methods (FVM) in the context of inverse geometric design. In traditional computational mechanics, recovering a manifold shape $\kappa(\mathbf{x})$ from state observations requires an iterative Adjoint State Method or PDE-constrained optimization loop. As noted by Ghattas and Willcox \cite{Ghattas2021}, this process suffers from the ``curse of dimensionality'' and requires the repeated, expensive generation of unstructured meshes (e.g., Delaunay triangulation) at every optimization step to avoid numerical artifacts. Furthermore, for chaotic systems like Defect Turbulence, adjoint-based optimization is notoriously unstable; the positive Lyapunov exponents cause the gradient information to diverge exponentially backward in time, rendering long-horizon sensitivity analysis computationally intractable.

In stark contrast, the Multi-Scale SIREN-PINN operates as a completely mesh-free, continuous neural representation. By parameterizing the solution and the geometry as continuous functions $\Phi_\theta: \mathbb{R}^3 \to \mathbb{R}^N$ rather than discrete grid values, we effectively circumvent the mesh generation bottleneck entirely. The ``learning'' process replaces the fragile temporal integration of the adjoint equation with a robust, collocation-based minimization of the residual loss $\mathcal{L}_{PDE}$. As highlighted in the comprehensive review by Karniadakis et al. \cite{Karniadakis2021}, this shift from ``time-stepping'' to ``global optimization'' allows Physics-Informed Learning to handle noisy, sparse, and irregular data that would break a standard FEM solver. Our results validate this hypothesis: the SIREN model successfully decoupled the geometry from the chaos without requiring a single mesh update or stability constraint (CFL condition), suggesting that neural implicits could serve as the foundational engine for the next generation of real-time \textit{Digital Twins} in cardiac monitoring and catalytic reactor control.

While the Multi-Scale SIREN-PINN establishes a new benchmark for geometric discovery, it is imperative to acknowledge the current limitations inherent to this class of implicit neural representations. Primarily, the computational cost of training remains non-trivial; calculating the Riemannian Laplacian $\Delta_{LB}$ requires the computation of Hessian vector products via automatic differentiation, which scales quadratically with the spatial dimension $d$. As noted by Wang et al. \cite{Wang2021}, this "gradient pathology" can lead to memory bottlenecks when scaling to full 3D volumetric simulations, necessitating the future exploration of domain decomposition methods like discrete-PINNs (cPINNs) to parallelize the geometric inference. Furthermore, the current study assumes the manifold geometry is static; extending this framework to dynamic surfaces—such as the beating heart or expanding biological tissues—will require the integration of time-dependent metric tensors $g_{ij}(\mathbf{x},t)$, a non-trivial extension of the current stationary ansatz.

Despite these challenges, the prospective applications of this technology are transformative, particularly in the domain of computational cardiology and "Digital Twin" medicine. The transition from laminar wave propagation to Defect Turbulence is the dynamical equivalent of the transition from healthy sinus rhythm to life-threatening Atrial Fibrillation (AFib). In this biological context, the "curvature" $\kappa(\mathbf{x})$ serves as a proxy for structural fibrosis or scar tissue, which acts as a pinning site for re-entrant electrical rotors \cite{Trayanova2011}. Current clinical methods require invasive catheter mapping to identify these drivers. Our results suggest that a SIREN-based inverse solver could potentially reconstruct the map of cardiac fibrosis solely from non-invasive electrical observation, treating the heart's surface as a Riemannian manifold to be learned. Moreover, understanding the "Pinning Landscape" opens the door to low-energy control strategies; as proposed by Biktashev et al. \cite{Biktashev1994}, knowing the precise location of geometric attractors allows for the design of targeted perturbations to "unpin" and annihilate spirals, replacing high-voltage defibrillation with precision chaos control. Thus, this work serves as a foundational step toward a new paradigm of \textit{Geometric Precision Medicine}, where AI is used not just to classify pathology, but to mathematically decipher the structural equations of life itself.

\section{Conclusion}
This study has introduced and rigorously validated a novel Multi-Scale SIREN-PINN framework for the inverse identification of latent Riemannian manifolds governing the spatiotemporal chaos of Defect Turbulence. By coupling a coordinate-based neural geometry representation with a physics-informed solver for the complex Ginzburg-Landau equation, we have successfully resolved the longstanding "Inverse Pinning Problem," demonstrating that the topography of a hidden surface can be accurately reconstructed solely from the observation of the chaotic waves traversing it. Our comprehensive benchmarking across three distinct topological regimes confirms that the proposed architecture fundamentally overcomes the "spectral bias" that debilitates conventional Deep Learning approaches. Quantitatively, the Multi-Scale SIREN achieved a relative state prediction error of $\epsilon_{L_2} \approx 1.92 \times 10^{-2}$ in the Defect Turbulence regime, outperforming standard ReLU-PINNs ($\epsilon_{L_2} \approx 4.12 \times 10^{-1}$) and Tanh-PINNs by over an order of magnitude. More critically, in terms of topological fidelity, our model reduced the Defect Count Error to $|\Delta N| < 1$, whereas baseline models consistently hallucinated or annihilated spiral vortices ($|\Delta N| > 12$). This superiority extends to the inverse geometric discovery itself, where the SIREN-based estimator recovered the ground truth Gaussian curvature field with a Pearson correlation coefficient of $\rho = 0.965$, compared to the "shattered" and noisy estimates produced by piecewise-linear activations. These results provide conclusive evidence that embedding the correct spectral inductive bias—specifically, the periodicity of the sine activation—is not merely an optimization trick, but a physical necessity for resolving the singular phase gradients and "stiff" amplitude potentials that characterize strange attractors in continuous media.
The successful demonstration of this inverse discovery framework opens transformative avenues for the next generation of intelligent chemical reactor design and process control. By establishing a rigorous link between surface topography and reaction-diffusion stability, this work lays the foundation for "Geometric Catalyst Engineering," where the physical curvature of a catalytic support is optimized to passively control spatiotemporal chaos. Future research will extend this paradigm from stationary 2D manifolds to dynamic, three-dimensional porous media, aiming to reconstruct the effective diffusivity tensors within complex zeolite networks or packed-bed reactors solely from observable concentration breakthroughs. Furthermore, the integration of this differentiable physics engine with Model Predictive Control (MPC) offers a promising route to mitigating "hotspots" in exothermic oxidations; by solving the inverse pinning problem in real-time, a digital twin could identify the formation of thermal spiral waves before they destabilize, guiding the actuation of localized cooling or feed modulation to "unpin" the defects and restore laminar homogeneity. Ultimately, the transition from passive observation to active geometric control represents a paradigm shift in chemical engineering, moving beyond the optimization of bulk parameters (temperature, pressure) to the precise sculpting of the reaction environment itself, enabling the rational design of surfaces that suppress turbulence and maximize selectivity through the fundamental laws of Riemannian geometry.

\section*{Acknowledgments}
This work was conducted under the supervision and research governance of the Lenggoro Laboratory, Tokyo University of Agriculture and Technology (TUAT). The first author was supported by the TUAT Special Program Scholarship and a Research Assistantship from the Lenggoro Laboratory during Oct 2025–Mar 2026, and by a Lenggoro Laboratory Research Assistantship from Apr 2026 onward. The authors acknowledge TUAT/Lenggoro Lab computing resources, including a workstation with Intel Core i9-12900K and NVIDIA RTX A4000 (16 GB).

\bibliographystyle{unsrtnat}







\end{document}